\documentclass[numberedappendix]{emulateapj}
\usepackage{rotating}
\usepackage{amsmath, color,graphicx,tablefootnote,subfigure}
\usepackage{graphics,subfigure,hyperref,float}
\usepackage[rightcaption]{sidecap}
\newcommand\ii{{\sc ii}}
\newcommand\iii{{\sc iii}}

\newcommand{\CH}[1]{\colhead{#1}}

\newcommand\oiii{[\ion{O}{3}]}
\newcommand\oii{[\ion{O}{2}]}
\newcommand\siii{[\ion{S}{3}]}
\newcommand\nii{[\ion{N}{2}]}
\newcommand\sii{[\ion{S}{2}]}
\newcommand\W{$\lambda$}

\submitted{}
\shorttitle{CHAOS VI}
\shortauthors{Rogers et al.}
\accepted{to ApJ on April 7th, 2021}

\usepackage{natbib}
\begin{document}

\title{CHAOS VI: DIRECT ABUNDANCES IN NGC 2403}

\author{Noah S. J. Rogers$^{1}$, 
			 Evan D. Skillman$^{1}$, 
			 Richard W. Pogge$^{2,3}$,
			 Danielle A. Berg$^{4}$, 
			 John Moustakas$^{5}$,
			 Kevin V. Croxall$^{6}$, 
			 Jiayi Sun$^{2}$}

\affiliation{
				 $^{1}$Minnesota Institute for Astrophysics, University of Minnesota, 116 Church St. SE, Minneapolis, MN 55455; \\
				 $^{2}$Department of Astronomy, The Ohio State University, 140 W 18th Ave., Columbus, OH, 43210; \\
				 $^{3}$Center for Cosmology \& AstroParticle Physics, The Ohio State University, 191 West Woodruff Avenue, Columbus, OH 43210; \\
				 $^{4}$Department of Astronomy, University of Texas at Austin, 2515 Speedway, Austin, Texas, 78712; \\
				 $^{5}$Department of Physics \& Astronomy, Siena College, 515 Loudon Road, Loudonville, NY 12211; \\
				 $^{6}$Expeed Software, 100 W Old Wilson Bridge Rd Suite 216, Worthington, OH 43085 \\
}

%\accepted{to \apj}

\begin{abstract}

We report the direct abundances for the galaxy NGC\,2403 as observed by the CHemical Abundances Of Spirals (CHAOS) project. Using the Multi-Object Double Spectrograph on the Large Binocular Telescope, we observe two fields with H\ii\ regions that cover an R$_g$/R$_{e}$ range of 0.18 to 2.31. 32 H\ii\ regions contain at least one auroral line detection, and we detect a total of 122 temperature-sensitive auroral lines. Here, for the first time, we use the intrinsic scatter in the T$_e$-T$_e$ diagrams, added in quadrature to the uncertainty on the measured temperature, to determine the uncertainty on an electron temperature inferred for one ionization zone from a measurement in a different ionization zone. We then use all available temperature data within an H\ii\ region to obtain a weighted average temperature within each ionization zone. We re-derive the oxygen abundances of all CHAOS galaxies using this new temperature prioritization method, and we find that the gradients are consistent with the results of \citet{berg2019}. For NGC\,2403, we measure a direct oxygen abundance gradient of $-$0.09($\pm$0.03) dex/R$_e$, with an intrinsic dispersion of 0.037($\pm$0.017) dex, and an N/O abundance gradient of $-$0.17($\pm$0.03) dex/R$_e$ with an intrinsic dispersion of 0.060($\pm$0.018) dex. For direct comparison, we use the line intensities from the study of NGC\,2403 by \citet{berg2013}, and find their recomputed values for the O/H and N/O gradients are consistent with ours.

\end{abstract}

%%%%%%%%%%%%%%%%%%%%%%%%%

\section{Introduction} \label{sec:intro}
Stellar nucleosynthesis enriches star-forming galaxies with heavy elements which are incorporated into the next generation of stars. Mapping the distribution of chemical abundances in galaxies gives insight into their stellar and chemical evolution, the yields of stellar nucleosynthesis, and the underlying physics of the interstellar medium (ISM). Beyond their use in individual galaxies, abundances are crucial for studies of the mass-metallicity relation (MZR) for galaxies \citep[e.g.,][]{lequ1979,trem2004} and as input parameters for other galaxy-related studies and models, like those concerning the CO-to-H$_{2}$ conversion factor \citep{sand2013}.

Optical emission lines from H\ii\ regions are the primary mechanism for obtaining chemical abundances in nearby spiral galaxies. At the typical temperature of an H\ii\ region (on the order of 10$^{4}$ K), the dominant cooling mechanism is the emission from collisionally-excited lines in metal ions like O$^{+}$, O$^{2+}$, N$^{+}$, S$^{+}$, and S$^{2+}$. Free electrons excite the outer electrons in these metal ions via collisions. The subsequent radiative de-excitations produce photons that will escape the region without exciting a similar energy level due to the scarcity of these metals relative to H$^+$. The ratios of specific forbidden line fluxes from these photons are used to directly calculate the electron temperatures and/or densities within the region. The temperatures and densities are used to calculate the emissivities of various transitions, which are applied in conjunction with the corresponding line fluxes to obtain the relative ionic abundances.

The auroral lines needed for this abundance technique, called the ``direct" abundance determination \citep{dine1990}, are faint, especially in regions of high metallicity/ low temperature where emission from collisionally-excited fine structure lines in the IR dominate the cooling. The intrinsic faintness of auroral lines has resulted in a lack of detections and, subsequently, in direct abundance studies with a relatively small number of H\ii\ regions \citep[e.g.,][and references therein]{berg2013}. Other abundance studies have developed calibrators relating the flux of the strongest emission lines to the chemical abundances within an H\ii\ region \citep[][and others]{page1979,mcGa1991,kobu1999,KD02, pett2004,bres2007,mari2013,pily2016}. However, comparing strong-line calibrators of different origin reveals differences in the inferred abundances. Empirical and theoretical strong-line abundances for the same galaxy produce discrepancies (sometimes sizeable, $\sim$0.7 dex at the extremes) in abundances, gradients, and dispersions \citep{KE08,mous2010}. Nearby galaxies present the best chance of observing the faint, temperature-sensitive auroral lines necessary for direct abundances. Increasing the sample of direct abundance measurements is necessary to move away from the use of strong-line calibrators in these galaxies, to create more statistically significant direct abundance studies, and to better calibrate empirical strong-line diagnostics for use in more distant galaxies.

Another method to obtain direct abundances involves stacking the spectra of many galaxies of similar stellar mass \citep{lian2007}, stellar mass and star-formation rate \citep{andr2013,brow2016}, or \oii\ and \oiii\ nebular line flux \citep{curt2017}. The assumption in doing so is that galaxies of similar stellar mass have similar metallicities by the MZR, or that galaxies containing similar strong-line emission have comparable O/H abundance. Given the faintness of the auroral lines and the difficulty of observing them at high redshift, stacking spectra allows for increased detections where traditional, single-object spectroscopy may otherwise fail to obtain a direct temperature. This is especially useful for establishing electron temperature trends \citep{andr2013} and for developing strong-line calibrators \citep{curt2017} using a statistically significant sample of galaxies.

Alternatively, Integral Field Unit (IFU) surveys observe the entire disk of a galaxy with many optical fibers. These surveys trade the high sensitivity needed to observe the temperature-sensitive auroral lines for complete spectral coverage of the galaxy. The result is strong-line detections in hundreds or thousands of pixels across the disk of the galaxy, allowing for statistically significant strong-line abundance analyses for many galaxies. For example, CALIFA and VLT/MUSE IFU surveys have identified a universal oxygen abundance gradient for 306 and 102 low-redshift galaxies, respectively \citep{sanc2014,same2018}. A spaxel-by-spaxel analysis for 122 galaxies in the CALIFA sample confirms this finding, although the universal gradient is slightly shallower \citep{same2016}. Additionally, \citet{ho2019} detected the \nii$\lambda$5755 auroral line in 80 H\ii\ regions in NGC\,1672 using VLT/MUSE, allowing for direct abundance determination in this galaxy. However, the wavelength coverage of IFU surveys often excludes some auroral or nebular emission lines that may be useful in determining temperatures of different ionization zones and direct abundances of different ionic species.

The CHemical Abundance Of Spirals (CHAOS) project has acquired high resolution observations of H\ii\ regions in nearby, face-on spiral galaxies from the \textit{Spitzer} Infrared Nearby Galaxies Survey \citep[SINGS;][]{kenn2003a} to increase the number of temperature-sensitive auroral line detections for use in abundance determinations. For a more detailed description of the CHAOS sample, see \citet{berg2015}. The CHAOS auroral and nebular line observations allow for temperature and abundance determination from multiple ions spanning a wide range of ionization. For example, \citet{berg2019} used the 180+ H\ii\ regions with two or more auroral line detections to find new T$_{e}$-T$_{e}$ relations, a possible explanation for the large intrinsic scatter within some direct abundance gradients, and the presence of a universal N/O gradient at R$_{g}$/R$_{e}$ $<$ 2.0. CHAOS has also detected a number of \ion{C}{2} recombination lines, resulting in a C/H abundance gradient for NGC\,5457 \citep{skil2020}. The CHAOS sample, once fully analyzed, can be used to find global trends in the direct abundances of nearby spiral galaxies, derive new ionization correction factors (ICFs) for unobserved ions within H\ii\ regions, study the discrepancy between direct and recombination line abundances, and to further address how the scatter in direct abundances is related to the physical processes of the ISM.

Using its large sample of direct abundances from H\ii\ regions, the CHAOS project has found significant variations amongst the intrinsic dispersions about the galactic oxygen abundance gradients. The magnitude of the dispersion in log(O/H) can be on the order of 0.1 dex, similar to the scatter \citet{roso2008} observed in the M33 H\ii\ regions when using the direct abundance method. Abundance dispersions, or lack thereof, are related to the physical processes within the ISM such as radial mixing, gas infall, or star formation \citep{roy1995}. Thus far, no uniform, statistically significant dataset of direct abundances in multiple galaxies has been capable of studying the abundance dispersions and how these are related to ISM properties, location within the spiral galaxy, or to galactic properties. IFU studies have detected abundance enhancements in spiral-arm H\ii\ regions relative to interarm regions \citep{ho2019,same2020}, though the magnitude of these variations is not well established \citep[see][]{krec2019}. The strong-line methods are used to measure these variations, and these methods produce smaller variations in oxygen abundance relative to the direct method \citep{arel2016}. CHAOS targets H\ii\ regions spanning the disk of each observed galaxy and can detect multiple temperature-sensitive auroral lines within a given region. This combination makes the total CHAOS database optimal for a study of the direct abundance dispersion in spiral galaxies.

NGC\,2403 is a nearby \citep[adopted distance of 3.18 Mpc from][]{tull2013}, intermediate spiral galaxy (SABcd) with $R_{25}$ of 10.95\arcmin\ and inclination of 63$^\circ$ \citep{blok2008}. This galaxy has been the focus of multiple abundance studies, but these studies have used the spectra from a relatively small number of H\ii\ regions. For example, \citet{garn1997} reported a direct abundance gradient in NGC\,2403 using the optical spectra of 12 H\ii\ regions, while \citet{berg2013} used the optical spectra of 7 H\ii\ regions to measure the abundance gradient. Recently, \citet{mao2018} conducted a spectroscopic study on 11, spatially resolved H\ii\ regions in NGC\,2403. They used commonly employed strong-line abundance calibrators to check for variations of the inferred abundances as a function of nebular radius. Absolute abundance discrepancies, in addition to different radial/ionization dependences, are found between different strong-line calibrators. Additionally, blue supergiant spectra have been acquired in NGC\,2403 (Bresolin, private communication). These spectra allow for a completely independent measure of the abundances in the galaxy.

Here, we present the results of the CHAOS observations of NGC\,2403, including the chemical abundances, gradients, and dispersions within the galaxy as found from the H\ii\ regions with detected temperature-sensitive auroral lines. The remainder of this paper is organized as follows: observations and data reduction are reported in \S2; \S3 highlights the electron temperature relations found in five CHAOS galaxies and presents a new technique to better estimate temperature uncertainties when applying empirical T$_{e}$-T$_{e}$ relations; in \S4 we describe how the abundances of each ionic species are determined; in \S5 we report the direct abundances gradients observed in NGC\,2403, compare these to updated literature values, and examine the abundances in the previous four CHAOS galaxies; \S6 details the $\alpha$ element abundance trends found in the present sample of H\ii\ regions; we summarize our findings in \S7.

%%%%%%%%%%%%%%%%%%%%%%%%%

\section{NGC 2403 Observations and Data Reduction}
\subsection{Data Acquisition} 

\begin{deluxetable}{lcccccccccc}  
\tabletypesize{\scriptsize}
\tablecaption{Adopted Properties of NGC\,2403}
\tablewidth{0pt}
\tablehead{ 
  \colhead{Property}	&
  \colhead{Adopted Value}	&
  \colhead{Reference}	
  }
\startdata
R.A. 	& 07$^h$36$^m$51.4$^s$ &  1 \\  
Dec  & +65$^\circ$36$^m$09.2$^s$ & 1 \\
Inclination & 63$^\circ$ & 2\\
Position Angle  & 124$^\circ$ & 2\\
Distance & 3.18$\pm0.12$ Mpc & 3 \\
log(M$_\star$/M$_\odot$) & 9.57 & 4\\
R$_{25}$ & 10.95\arcmin & 5 \\
R$_e$ & 178.0$\pm$5\arcsec & 6 \\
Redshift & 0.000445 & 1
\enddata
\label{t:n2403global}
\tablecomments{Units of right ascension are hours, minutes, and seconds, and units of declination are degrees, arcminutes, and arcseconds. References are as follows: [1]  2MASS Extended Objects Final Release  [2] \citet{blok2008} [3] \citet{tull2013} [4] \citet{lero2019} [5] \citet{kend2011} [6] The effective radius is calculated in the manner described in Appendix C of \citet{berg2019}}
\end{deluxetable}

The Multi-Object Double Spectrograph \citep[MODS,][]{pogg2010} on the Large Binocular Telescope \citep[LBT,][]{hill2010} is used to observe two fields of NGC\,2403. Custom laser-cut slit masks allow for the concurrent observation of $\sim$20 objects per field, which takes advantage of multiplexing and makes the data less susceptible to nightly differences in atmospheric variations. MODS uses the G400L grating (400 lines mm$^{-1}$; R$\approx$1850) for the blue and the G670L grating (250 lines mm$^{-1}$; R$\approx$2300) for the red to cover a wavelength range of 3200 – 10000 \AA, appropriate for the optical auroral lines necessary for temperature, density, and abundance determination. The slits are cut to be 1\arcsec\ in width and can range from 10\arcsec\ to 32\arcsec\ in length. The field of view of MODS is 6\arcmin $\times$6\arcmin, roughly half the $R_{25}$ of NGC\,2403 but large enough to achieve coverage of the disk H\ii\ regions in two fields. The resolution of MODS allows for auroral line detections even for lines that are near other nebular emission lines (for instance, \siii$\lambda$6312 and [\ion{O}{1}]$\lambda$6300). As a result, MODS and the LBT are optimized for optical direct abundance studies of nearby spiral galaxies.

The first and second fields were observed on November 16th, 2017 and February 8th, 2018, respectively. Both fields were simultaneously observed in the blue (3200 - 5800 \AA) and red (5500 - 10000 \AA) with MODS1, the detector on the LBT SX telescope. The seeing during the nights of observation range from 1.0\arcsec\ to 1.3\arcsec. The fields were observed for 6 exposures of 1200 seconds each, at an air mass between 1.2 and 1.4, and at a position angle near the parallactic angle halfway through observation. Observing at low air mass and an optimal parallactic angle, we ensure minimal loss of intensity at the ends of the observed spectrum due to differential atmospheric refraction \citep{filippenko1982}.

\begin{figure*}[!t] 
\epsscale{1}
   \centering
   \plotone{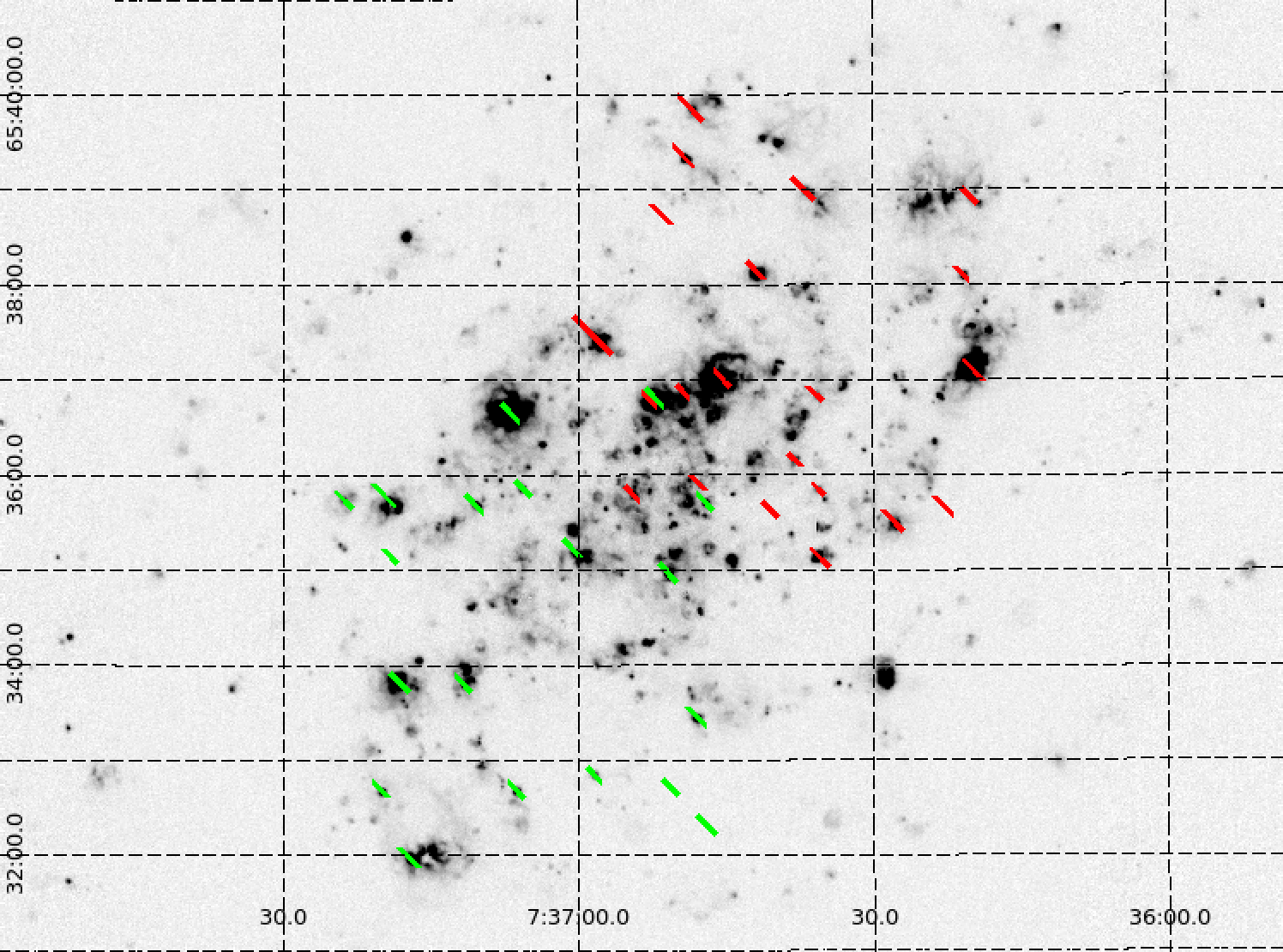}
   \caption{Map of the slits used to observe the H\ii\ regions in NGC\,2403 overlaid on a continuum-subtracted H$\alpha$ image (obtained from http://sumac.astro.indiana.edu/$\sim$vanzee/LVL/NGC2403/). The axes are in degrees: RA on x axis and Dec on y axis. The slits are color-coded by observation field: Green for Field 1, Red for Field 2.\\}
   \label{fig:halphaimg}
\end{figure*} 

Slits are cut to target bright H\ii\ regions spanning the disk of the galaxy. The slit locations are shown in Figure \ref{fig:halphaimg}, superimposed on a continuum-subtracted H$\alpha$ image of NGC\,2403\footnote{from Liese van Zee, http://sumac.astro.indiana.edu/$\sim$vanzee/ LVL/NGC2403/} and color coded by field. The targeted H\ii\ regions range in R$_{g}$/R$_{e}$ from 0.18 to 2.31, sufficient to examine the radial dependence of the chemical abundances. In this paper, the IDs of the targets are provided in terms of the right ascension and declination offsets of their surface brightness peaks relative to the center of NGC\,2403, as given in Table \ref{t:n2403global}. Table \ref{t:locations} provides the radial locations of, and the auroral lines detected in, each of the 33 H\ii\ regions targeted. Additionally, the last column of Table \ref{t:locations} notes the presence of Wolf-Rayet (WR) features in the optical spectra. WR stars are high-mass stars with strong stellar winds and are characterized, spectrally, by broadened emission features from \ion{He}{2}, \ion{C}{3}, \ion{C}{4}, and \ion{N}{3} \citep{lope2010}. These features are typically blended together at $\sim$4600 - 4700 \AA\ (the blue WR bump) and $\sim$5750 - 5870 \AA\ (the red WR bump). The blue WR bump is the most common WR feature observed in the H\ii\ regions of NGC\,2403, and a blue WR feature always accompanies a red WR feature when the latter is observed in an H\ii\ region (8 objects total). Table \ref{t:locations} also provides the regions which contain \ion{C}{2} $\lambda$4267 \AA\ detections. This recombination line allows for the determination of C$^{++}$ abundances \citep[for example,][]{este2009,este2020,skil2020}, but we reserve a study of the recombination line abundances in multiple galaxies for a future study.

\subsection{MODS Data Reduction and Line Modeling}

\begin{figure*}[pt]
\epsscale{1.0}
   \centering
   \plotone{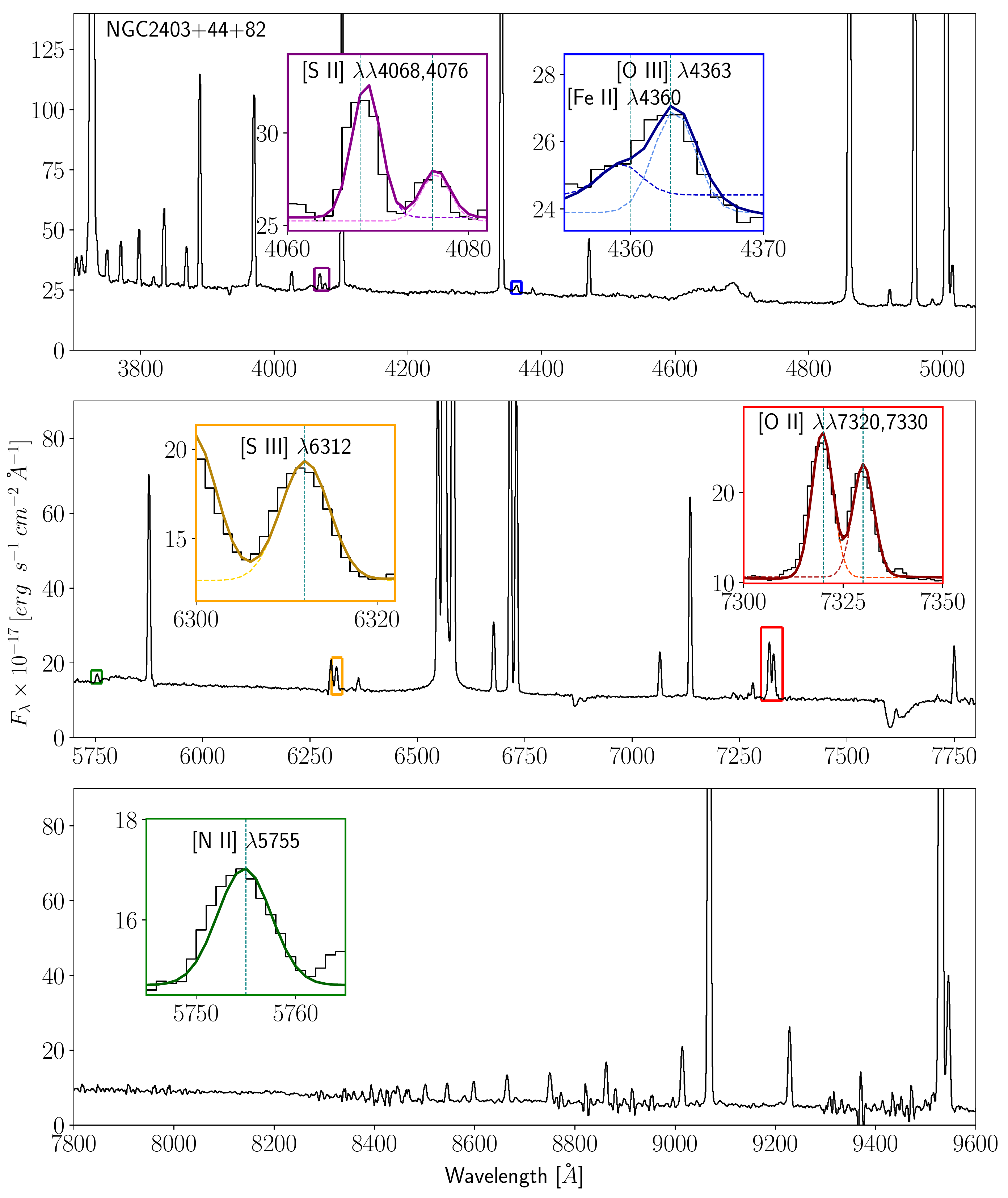}
   \caption{A 1D flux- and wavelength-calibrated spectrum from MODS observations of the region NGC\,2403+44+82. Faint, temperature-sensitive auroral lines necessary for direct abundance calculation are magnified in the subplots. The Gaussian fits to the auroral lines and the total fit to the spectra are represented by dashed lines and bold lines, respectively, within each subplot. Additionally, the fit to [\ion{Fe}{2}]$\lambda$4360 is provided. In this spectrum, the \sii$\lambda$\,$\lambda$4068,4076, \oiii$\lambda$4363, \nii$\lambda$5755, \siii$\lambda$6312, and \oii$\lambda\lambda$7320,7330 lines are all detected at S/N $\geq$ 3. The Wolf-Rayet features are also present between 4600 and 4800 \AA, and between 5750 and 5900 \AA.\\}
   \label{fig:spectra}
\end{figure*}

We highlight the key steps of the CHAOS project's data reduction pipeline\footnote{The MODS reduction pipeline was developed by Kevin Croxall with funding from NSF Grant AST-1108693.\ Details at http://www.astronomy.ohio-state.edu/MODS/Software/modsIDL/} briefly. For a complete explanation of the MODS data reduction pipeline, the reader is referred to \citet{berg2015}. The modsCCDRed \textsc{Python} programs are used to bias subtract and flat-field the raw CCD images of the science targets, standard stars, and calibration lamps. The resulting images are used in the current version of the modsIDL reduction pipeline, which runs in the XIDL reduction package\footnote{http://www.ucolick.org/$\sim$xavier/IDL/}. Sky subtraction and region extraction is performed on each slit, and calibration lamp data provides a wavelength calibration for the resulting 1D spectrum. Standard stars are used for flux calibration and correction for atmospheric extinction \citep{bohlin2014}. There are three sky-only slits (skyslits) in each of the field masks (see Figure \ref{fig:halphaimg}). In some slits, selecting a region for sky subtraction is not possible without additionally removing a substantial amount of flux from the H\ii\ region. This is the case for H\ii\ regions larger than the length of the slit, such as the large regions NGC\,2403$-$38+51, NGC\,2403+7+37, \& NGC2403+96+30. For slits where local sky is unobtainable, the fitted sky spectrum from one of these skyslits, or a neighboring slit with ample sky, is used for sky subtraction. Figure \ref{fig:spectra} shows an example of a 1D, flux- and wavelength-calibrated spectrum from one of the H\ii\ regions (NGC\,2403+44+82) containing five auroral line detections.

The spectrum of an H\ii\ region contains the underlying stellar continuum from the stars ionizing the region. The STARLIGHTv04\footnote{www.starlight.ufsc.br} spectral synthesis code \citep{starlight} is used in conjunction with the stellar population models of \citet{bruz2003} to model the stellar continuum. We do not constrain the metallicity of the input stellar models because the shape of the stellar continuum is only used for the continuum component of line-fitting code. We do not use the modeled underlying absorption features for reddening corrections (see \S2.3). After subtracting the stellar continuum from the spectrum, each emission line is fit with a Gaussian profile while allowing for an additional nebular continuum component. The Gaussian fits work well for most lines, including the auroral lines needed to determine the electron temperatures and densities.

As discussed in \citet{berg2019}, the \oii$\lambda\lambda$3726, 3729 doublet, a density-dependent set of emission lines, is blended at the resolution of MODS (see Figure \ref{fig:spectra}). We use the FWHM of the neighboring lines and the wavelength separation of the two transitions to simultaneously fit the two lines. The total flux in the doublet is reported as the flux of “\oii$\lambda$3727”. Additionally, [\ion{Fe}{2}]$\lambda$4360 can contaminate \oiii$\lambda$4363 at high metallicities (12+log(O/H) $\geq$ 8.3) where this line may otherwise be weak or undetected \citep{curt2017}. If [\ion{Fe}{2}]$\lambda$4360 is misinterpreted as a detection of \oiii$\lambda$4363, or if the flux of [\ion{Fe}{2}]$\lambda$4360 is blended with the flux of \oiii$\lambda$4363, then temperatures (abundances) derived for that region are biased unphysically high (low). \citet{berg2019} found evidence for such contamination in a small number of H\ii\ regions in the previous CHAOS galaxies. As such, [\ion{Fe}{2}]$\lambda$4360 and \oiii$\lambda$4363 are fit simultaneously to avoid contamination.

The MODS spectra have a loss in sensitivity near the dichroic wavelength cutoff at 5700 \AA. This can introduce non-physical features around the wavelength crossover, which makes the stellar continuum difficult to fit properly and can result in missed \nii$\lambda$5755 detections. To correct for this, the stellar continuum across the dichroic is fit with a low-order polynomial. Then, these features in the blue and red continuum are fit with high-order polynomials, taking precaution to avoid any emission lines and WR features. With the continuum and the features modeled by their respective polynomials, the difference between the polynomials is taken and applied to the blue and red spectra. The net result is that the blue and red spectra now match at the dichroic and that the spectra follow the stellar continuum across the dichroic, enabling a better fit for \nii$\lambda$5755.

Electron temperatures are exponentially sensitive to the auroral-to-nebular line ratios. Therefore, accurate flux measurements, particularly of the weak auroral lines, are essential to the electron temperature determination. While the fitting program used in the MODS data reduction pipeline is able to easily fit the strongest lines, the auroral lines require more care. This is particularly true when night sky lines close to the auroral lines (e.g., \ion{Hg}{1} $\lambda$4358 \AA) show significant subtraction residuals, or if the modeled stellar continuum is not well fit due to differences in the stellar absorption features or other noise structures. Consistent with all CHAOS galaxies, each auroral line is fit by hand and the results compared with those from the MODS data reduction pipeline. In instances where the modeled stellar continuum or night sky noise impact the line fit from the reduction pipeline, the line fluxes and RMS in the continuum are updated to be that of the hand fits. The pipeline fits are used in all other cases to maintain consistency with the fitted strong lines.

\begin{deluxetable*}{lccccc|c|c|c|c|c|c|c}
	\tabletypesize{\scriptsize}
	\tablecaption{NGC\,2403 MODS/LBT Observations}
	\tablewidth{0pt}
	\tablehead{   
	  \colhead{H~\ii}	&
	  \colhead{R.A.}		&
	  \colhead{Dec.}	&
	  \colhead{$R_{g}$}  &
	  \colhead{$\frac{R_g}{R_{e}}$}	&
	  \colhead{$R_{g}$}  &
	  \multicolumn{5}{c}{Auroral Line Detections}	&
	  \colhead{RLs} &
	  \colhead{Wolf}  \\
	  \colhead{Region}	&
	  \colhead{(2000)}	&
	  \colhead{(2000)}	&
	  \colhead{(arcsec)}	&
	  \colhead{}	&
	  \colhead{(kpc)}		&
	  \colhead{[O\,\iii]}	&
	  \colhead{[N\,\ii]}	&
	  \colhead{[S\,\iii]}	&
	  \colhead{[O\,\ii]}	&
	  \colhead{[S\,\ii]}    &
	  \colhead{C\,\ii}		&
	  \colhead{Rayet}
	  }
\startdata
Total Detections: & & & & & & 18 & 18 & 28 & 32 & 26 & 5 & 13 \\ \hline
NGC2403+19$-$22	&	7:36:54.4	&	65:35:47.24	&	32.58	&	0.18	&	0.50	&	 &		&		&	\checkmark	&  &	 &	  \\
NGC2403$-$23$-$16	&	7:36:47.6	&	65:35:52.93	&	59.03	&	0.33	&	0.91	&	 &		&		&	 &		&	   &   \\
NGC2403$-$14+42	&	7:36:49.2	&	65:36:51.10	&	69.18	&	0.39	&	1.07	&	 &	\checkmark	&		\checkmark &  \checkmark		&  \checkmark & 	&	 \checkmark \\
NGC2403$-$38+51	&	7:36:45.2	&	65:36:59.95	&	75.43	&	0.42	&	1.16	& 	\checkmark &	\checkmark	&	\checkmark  	&   \checkmark	  & \checkmark & 		\checkmark &	 \checkmark \\
NGC2403+7+37	&	7:36:52.6	&	65:36:45.93	&	77.57	&	0.44 &	1.20	&	\checkmark &	\checkmark	& \checkmark	  	&   \checkmark	  & \checkmark & \checkmark	&	 \checkmark \\
NGC2403$-$27$-$28	&	7:36:46.9	&	65:35:41.04	&	85.37	&	0.48	&	1.32	&	 &	 	& \checkmark	&   \checkmark & \checkmark	&	&	  \\
NGC2403+56$-$59	&	7:37:00.5	&	65:35:10.39	&	88.20	&	0.50	&	1.36 &	 &		&	& \checkmark	&  &	  &	 \\
NGC2403+88$-$18	&	7:37:05.5	&	65:35:50.85	&	111.75	&	0.63	&	1.72 &	 &		&	\checkmark	&	\checkmark	& \checkmark &	 	&	\checkmark \\
NGC2403$-$97+39	&	7:36:35.6	&	65:36:48.48	&	113.24	&	0.64	&	1.75	&	 &		&	\checkmark 	& 	\checkmark  &  &	  & 	\\
NGC2403$-$84$-$0	&	7:36:37.8	&	65:36:09.07	&	124.50 &	0.70	&	1.92	&	 &		&	\checkmark  	& \checkmark  &  &	 & 	\\
NGC2403$-$3$-$71	&	7:36:50.8	&	65:34:58.60	&	138.04	&	0.78	&	2.12	&	 &		&	\checkmark  	& \checkmark  &  &	 & 	\\
NGC2403+119$-$28	&	7:37:10.5	&	65:35:41.61	&	148.62	&	0.83	&	2.29	&\checkmark	 &	\checkmark	&	\checkmark  	& \checkmark    & \checkmark &  &	 \checkmark	\\
NGC2403$-$98$-$19 &	7:36:35.5	&	65:35:49.86	&	171.71	&	0.96	&	2.65 &	 &	 &	\checkmark &	\checkmark & \checkmark &  &	  	\\
NGC2403$-$59+118	&	7:36:41.8	&	65:38:07.23	&	183.34	&	1.03	&	2.83	&	\checkmark &	\checkmark	& \checkmark	  	&   \checkmark &  \checkmark & &	 \checkmark	\\
NGC2403+96+30	&	7:37:06.8	&	65:36:39.25 &	183.51	&	1.03	&	2.83	&\checkmark	 &	\checkmark	&	\checkmark  	& \checkmark    & \checkmark &	 \checkmark & \checkmark	\\
NGC2403+44+82 &	7:36:58.5	&	65:37:31.55	&	205.28 &	1.15	&	3.16	&	\checkmark &	\checkmark	& \checkmark	  	&   \checkmark	  & \checkmark  & \checkmark	& \checkmark \\
NGC2403+125$-$142 &	7:37:11.6	&	65:33:46.66 &	211.84	&	1.19	&	3.27	&	\checkmark &	& \checkmark	  	&   \checkmark	  & \checkmark  &  &	 	\\
NGC2403+166$-$140 &	7:37:18.1	&	65:33:49.15	&	221.78	&	1.25	&	3.42	&	\checkmark &	\checkmark	& 	\checkmark  	&   \checkmark	  & \checkmark  &  &	  \checkmark	\\
NGC2403$-$190+116 &	7:36:20.6	&	65:38:04.55	&	223.55	&	1.26	&	3.45	&	\checkmark &	\checkmark	& 	\checkmark  	&   \checkmark	  & \checkmark  &  &	 	\\
NGC2403+174$-$24 &	7:37:19.5	&	65:35:45.51	&	233.39	&	1.31	&	3.60	&	\checkmark &	\checkmark	& 	\checkmark  	&   \checkmark	  & \checkmark  &  &	 	\\
NGC2403$-$99$-$59 &	7:36:35.4	&	65:35:09.72	&	234.98	&	1.32	&	3.62	&	\checkmark &	\checkmark	& \checkmark	  	&   \checkmark	  & \checkmark &  &	  \checkmark	\\
NGC2403$-$196+58 &	7:36:19.7	&	65:37:07.49	&	237.12 &	1.33	&	3.66	&	\checkmark &	\checkmark	& 	\checkmark  	&   \checkmark	  & \checkmark &  &	  \checkmark	\\
NGC2403$-$194+165 &	7:36:19.9	&	65:38:53.91	&	260.70	&	1.46	&	4.02	&	 &		& 	\checkmark  	&   \checkmark	  &   &	 & 	\\
NGC2403$-$89+171	&	7:36:37.0	&	65:39:00.61	&	264.96	&	1.49	&	4.08 &	 &		&	\checkmark	&	\checkmark	&  \checkmark &	 &  	\\
NGC2403$-$146$-$38 &	7:36:27.8	&	65:35:30.96	&	268.18	&	1.51	&	4.13 &	 &	\checkmark	& \checkmark	& \checkmark	& \checkmark  &	  & \checkmark 	\\
NGC2403+201$-$24	&	7:37:23.8	&	65:35:44.59	&	271.97	&	1.53	&	4.19 &	 &		&		&	\checkmark	& \checkmark	&	 & 	 \\
NGC2403+178$-$210	&	7:37:19.9	&	65:32:38.70 &	312.22	&	1.75	&	4.81	&	\checkmark &		&	\checkmark  	& \checkmark    & \checkmark  &	 & 	\\
NGC2403$-$22$-$162 &	7:36:47.8	&	65:33:26.94	&	331.31	&	1.86	&	5.11	&	\checkmark &	\checkmark	& \checkmark	  	&   \checkmark	  & \checkmark &  &	  \checkmark	\\
NGC2403+92$-$210 &	7:37:06.2	&	65:32:39.11	&	332.45	&	1.87	&	5.13	&	 &	\checkmark	& \checkmark	  	&   \checkmark	  & \checkmark &	  & 	\\
NGC2403+43$-$200	&	7:36:58.3	&	65:32:49.65	&	344.41	&	1.93	&	5.31 &	\checkmark &	&	\checkmark  	& \checkmark    & \checkmark &	  & 	\\
NGC2403$-$14+192	&	7:36:49.1	&	65:39:21.19	&	353.85	&	1.99	&	5.46	&	\checkmark &	\checkmark	&		\checkmark &  \checkmark		&  \checkmark &	 &  	\\
NGC2403+160$-$251	&	7:37:17.1	&	65:31:57.98	&	378.04&	2.12	&	5.83	&	\checkmark &	\checkmark	&		\checkmark &  \checkmark		&  \checkmark  &	 \checkmark & 	\\
NGC2403$-$18+224	&	7:36:48.4	&	65:39:53.26	&	411.40	&	2.31	&	6.34	&	\checkmark &	\checkmark	&	\checkmark   &  \checkmark		&  \checkmark &	 &	
\enddata
	\label{t:locations}
	\tablecomments{H\ii\ regions observed in NGC\,2403 using MODS on the LBT.  The H\ii\ region ID, which is the offset in R.A. and Dec., in arcseconds, from the central position listed in Table \ref{t:n2403global}, is listed in Column 1. The right ascension and declination of the individual H\ii\ regions are given in units of hours, minutes, and seconds, and degrees, arcminutes, and arcseconds, respectively, in Columns 2 and 3. Radial distances of the regions are given in Columns 4 (in arcsec), 5 (normalized to R$_{e}$), and 6 (in kpc).  Columns 7-11 mark the regions that have [O\,\iii] $\lambda$4363, [N\,\ii] $\lambda$5755, [S\,\iii] $\lambda$6312, [O\,\ii]$\lambda\lambda$7320,7330, and [S\,\ii] $\lambda\lambda$4068,4076 auroral lines detections with S/N $>$ 3. If only one of the \oii\ or \sii\ lines is detected at S/N $>$ 3 then the region is still marked with a checkmark. Column 12 indicates a \ion{C}{2} $\lambda$4267 recombination line detection, and Column 13 denotes the presence of Wolf-Rayet features in the H\ii\ region spectrum.}
\end{deluxetable*}

Equation 2 from \citet{berg2013} approximates the uncertainty in the flux of an emission line. This equation is reproduced here:
\begin{equation}
\delta F_{\lambda} \approx \sqrt{(2\times\sqrt{n_{p}}\times\/RMS)^{2} + (0.02\times\/F_{\lambda})^{2}},
\end{equation}
where $n_{p}$ is the number of pixels over the FWHM of the line profile, RMS is the root mean squared noise in the continuum around the line, and $F_{\lambda}$ is the flux of the line. The pixel scale of MODS is 0.5 \AA\ per pixel, so the first term is simplified to $2\times \sqrt{2\times FWHM}\times RMS$. The uncertainty of weak lines is dominated by the RMS noise in the continuum about the fit, whereas the uncertainty of the strong lines is dominated by the 2\% uncertainty in the flux associated with the flux-calibration uncertainty when using standard stars \citep{oke1990}. We consider a line detected if its signal-to-noise ratio (S/N), or $F_{\lambda}/\delta F_{\lambda}$, is greater than 3.

\subsection{Reddening Corrections}

Line-of-sight reddening and the stellar absorption equivalent width are calculated in a manner similar to that described in \citet{oliv2001} with an MCMC component introduced by \citet{aver2011}. Recently, \citet{aver2020} introduced new calibrations using additional hydrogen recombination lines and the BPASS stellar evolution models \citep{eldr2009,eldr2017} to obtain the stellar absorption scaling coefficients. The observed hydrogen recombination line fluxes are affected by the underlying stellar absorption, $a_{H}$, or the equivalent width of the absorption feature, and reddening, $C(H\beta)$. Given $a_{H}$, $C(H\beta)$, and the electron temperature, T$_e$, we calculate theoretical Balmer line fluxes and compare these to the observed Balmer line fluxes. The combination of $a_{H}$ and $C(H\beta)$ that minimize the $\chi^2$ function \citep[see Equation A.2 in][]{aver2020} are chosen as the best-fit values. Uncertainties on $C(H\beta)$ are calculated by fixing $a_H$ to the best-fit value and generating a distribution of $C(H\beta)$ around its best-fit value. Using these inputs, a distribution of theoretical line fluxes is generated and used to repeat the $\chi^2$ minimization. The values of $C(H\beta)$ at which the $\chi^2$ function equals 1 (the 68\% confidence level) are averaged and taken as the uncertainties on $C(H\beta)$. The uncertainty on $a_H$ is calculated in a similar fashion, except that the lower uncertainty cannot be negative.

This method is similar to the method described in \citet{aver2011} and \citet{aver2020}, except the only parameters that are determined are $a_{H}$ and $C(H\beta)$. While additional free parameters and high-order hydrogen and helium recombination lines could be considered \citep[see][]{aver2020}, we use the flux ratios of H$\alpha$/H$\beta$, H$\gamma$/H$\beta$, and H$\delta$/H$\beta$ only. We choose to do this because the Paschen lines can be hard to detect in low surface brightness H\ii\ regions or when the sky subtraction is of low quality. Additionally, the fluxes of high-order Balmer lines above the stellar absorption features are difficult to accurately fit in a consistent and automatic manner. We calculate the electron temperature using the available auroral lines and fix the electron density to n$_{e}=10^{2}$ cm$^{-3}$ instead of leaving these as free parameters. 

To implement this method, a linear continuum is fit across the four most intense Balmer lines before stellar continuum subtraction. Although the modeled stellar continuum from STARLIGHT provides a fit to the Balmer absorption features, we leave these as free parameters by fitting a linear continuum around the chosen Balmer lines. The reddening law adopted is from \citet{card1989} using $A_{V}=3.1\times\/E(B-V)$, and the electron temperature and density are initially set to T$_{e}=10^{4}$ K and n$_{e}=10^{2}$ cm$^{-3}$, respectively. The theoretical line ratios relative to H$\beta$ for a given T$_e$ and n$_e$ are calculated using \textsc{PyNeb} \citep{luri2012,luri2015} in conjunction with the recombination line intensities from \citet{stor1995}. The $\chi^2$ function is minimized to find the values of $a_{H}$ and $C(H\beta)$ that best reproduce the observed Balmer line flux ratios. The observed fluxes of the Balmer lines are updated to account for the underlying stellar absorption, adopting the scaling coefficients reported in \citet{aver2020}. The errors on the Balmer lines are updated to account for the error in the flux of the line and the error on $a_H$. The spectrum is then de-reddened, and the error in the line flux and error on the reddening correction are combined in quadrature to obtain the reddening-corrected flux error. Electron temperatures are calculated from the available auroral lines in the de-reddened spectrum. Using the new electron temperature calculated in the high-ionization zone as the input electron temperature for the theoretical Balmer line calculation, we repeat this process until the change in the electron temperature is $<$ 20 K. If no auroral lines are detected in an H\ii\ region, the reddening correction is performed only once at T$_{e}=10^{4}$ K.

Table \ref{t:intensities} reports the line intensities relative to H$\beta$ measured from the H\ii\ regions in NGC\,2403. $a_{H}$, $C(H\beta)$, and the flux of H$\beta$ before accounting for the underlying stellar absorption are also included for each region. The Balmer line intensities typically agree with the expected theoretical ratio within statistical uncertainties, although there is trend that I(H$\gamma$)/I(H$\beta$) is systematically greater than the theoretical value. This approach to the reddening correction is unique to NGC\,2403 and has not been applied to the previously reported CHAOS galaxies. We intend to update the reddening corrections in those galaxies to be consistent with that of NGC\,2403, but doing so here is beyond the scope of this paper.

%%%%%%%%%%%%%%%%%%%%%%%%%%%%%
\section{Electron Temperatures}

The typical H\ii\ region spectrum in NGC\,2403 contains multiple temperature-sensitive auroral lines that span the ionization zones of an H\ii\ region. The three ratios we use to find the temperatures necessary for abundance analysis are $\mbox{[N II]}\lambda5755/\lambda\lambda6548,6584$, $\mbox{[S III]}\lambda6312/\lambda\lambda9069,9532$, and \oiii$\lambda$4363/$\lambda\lambda$4959,5007. It is also common to observe the temperature-sensitive lines \oii$\lambda\lambda$7320,7330 and \sii$\lambda\lambda$4069,4076 (see Figure \ref{t:locations}). However, we currently do not use the \oii\ or \sii\ auroral lines for abundance determination due to the scatter in the measured temperatures from these ions \citep[see discussion in \S3.2.4 and in][]{kenn2003b,croxall2016,berg2019}. The strong nebular components of the above ratios are easily detected, although water vapor absorption features can contaminate the far-red strong lines of \siii. The ratio of the \siii$\lambda$9532 and \siii$\lambda$9069 emissivities is constant over the range of temperatures and densities typical of an H\ii\ region. This ratio is j$_{\textrm{[S III]}9532}$/j$_{\textrm{[S III]}9069} =$ 2.47 as calculated using \textsc{PyNeb}. As such, the ratio of \siii$\lambda$9532 and \siii$\lambda$9069 fluxes is used as a diagnostic to determine if contamination has occurred in the far red. If I(\siii\/9532)/I(\siii\/9069) is within uncertainty of 2.47, then $I(\mbox{[S III]}6312)/(I(\mbox{[S III]}9532) + I(\mbox{[S III]}9069))$ is used to calculate the \siii\ temperature. If the intensity ratio is greater than 2.47, then $I(\mbox{[S III]}6312)/(I(\mbox{[S III]}9532)\times (1+2.47^{-1}))$ is used for T$_e$\siii\ determination, which corrects for an absorption-contaminated \siii$\lambda$9069 emission line. Similarly, if the intensity ratio is less than the theoretical value, as is the case for a \siii$\lambda$9532 that is too weak, the ratio $I(\mbox{[S III]}6312)/(I(\mbox{[S III]}9069)\times (1+2.47))$ determines T$_e$\siii.

\textsc{PyNeb} calculates the best-fit electron temperatures from the measured auroral-to-nebular line flux ratios described above, while the density-sensitive ratio \sii$\lambda$6716/\sii$\lambda$6731 determines the electron density. The electron density can be used in temperature determinations, but the emissivities of the auroral line transitions are nearly density independent in the low-density limit (n$_{e}\leq\/10^{3} $ cm$^{-3}$). The low-ionization zone electron temperature (see below) is used for density calculations. All regions in NGC\,2403 have densities in the low-density limit, so all temperatures are calculated at n$_{e}=10^{2}$ cm$^{-3}$. MCMC analysis is employed to determine the uncertainty on the electron temperatures: a range of possible flux ratios is generated using the measured flux ratio and its uncertainty, resulting in new temperatures for each generated ratio. The standard deviation of the temperature distribution is taken as the uncertainty on the best-fit temperature. Density uncertainties are calculated in the same manner.

As in previous CHAOS analyses, each H\ii\ region is split into three ionization zones. Auroral line emission from particular ions characterizes each zone. For example, emission from \nii$\lambda$5755 originates in the low-ionization zone while \oiii$\lambda$4363 is measured in the high-ionization zone. The low-ionization zone temperature is used when calculating the abundances of O$^{+}$, N$^+$, and S$^+$; the intermediate-ionization zone temperature is used for the abundances of S$^{2+}$ and Ar$^{2+}$; the high-ionization zone temperature is used for the abundances of O$^{2+}$ and Ne$^{2+}$.

\subsection{\texorpdfstring{T$_e$}{Te}-\texorpdfstring{T$_e$}{Te} Relation Methods}

Ideally, an H\ii\ region contains the necessary auroral lines to obtain direct temperatures for each ionization zone. This is not always the case, so T$_{e}$-T$_{e}$ relations are employed to infer the electron temperature in an ionization zone from the direct electron temperature in another zone. Previous studies have determined T$_{e}$-T$_{e}$ relations through photoionization models \citep[][and others]{camp1986,garn1992,page1992,izot2006,lope2012}, or empirically through auroral line detections in multiple ionization zones \citep[][]{este2009,pily2009,andr2013,croxall2016,yate2020,arel2020}. CHAOS provides an optimal dataset for the latter relations, as the electron temperature data from the numerous spiral galaxies span a large range in T$_{e}$ parameter space. For example, \citet{croxall2016} used the 70+ H\ii\ regions of NGC\,5457 with auroral line detections as a homogeneous dataset to develop linear, empirical T$_{e}$-T$_{e}$ relations for T$_{e}$\nii\/, T$_{e}$\siii\/, and T$_{e}$\oiii\ (see their Equations 5-7). \citet{berg2019} applied these relations to recalculate the abundances for all the CHAOS galaxies in a uniform manner.

The fit parameters for linear T$_{e}$-T$_{e}$ relations should be invertible such that differences in observational uncertainties associated with obtaining temperatures in different ionization zones do not bias the inferred temperatures. For example, a highly-ionized H\ii\ region might contain a well-measured \oiii\ temperature from the dominant ionization zone and a poorly measured \siii\ temperature from a physically smaller ionization zone. The opposite scenario is possible in an H\ii\ region dominated by the low- and intermediate-ionization zones. Electron temperatures in different ionization zones are dependent on similar properties (e.g., degree of ionization and metallicity). However, linear T$_e$-T$_e$ relations which assume that one temperature is an independent variable, coupled with the asymmetric uncertainties on the measured temperatures, can result in non-invertible linear fits, depending on the applied fitting technique.

Orthogonal Distance Regression (ODR) is used to obtain invertible, linear T$_e$-T$_e$ relations  while considering the errors on the observed temperatures. ODR-generated relations are invertible by construction, but the fitted intrinsic dispersion about the best-fit relation is dependent on which temperature is assumed to be the dependent variable. Here, the intrinsic dispersion is defined as random scatter in the dependent variable about the best-fit regression. The intrinsic dispersion about a best-fit T$_{e}$-T$_{e}$ relation can be interpreted as how much an ionization zone can deviate in temperature based on the H\ii\ region’s physical structure.  Obtaining a temperature in an ionization zone via a T$_{e}$-T$_{e}$ relation disregards true departures from the relationships, departures that a direct temperature may better represent. Therefore, it is critical to develop relations for each ionization zone that appropriately infer temperatures in the other zones, and that use the intrinsic dispersion about each relation to better account for the unique physical conditions in each ionization zone. In \S3.2 we describe the individual T$_e$-T$_e$ relationships found for the CHAOS data, and we discuss our new methodology for applying the relationships in \S3.3.

\begin{figure*}[t]
   %\epsscale{1}
   \centering
   \plotone{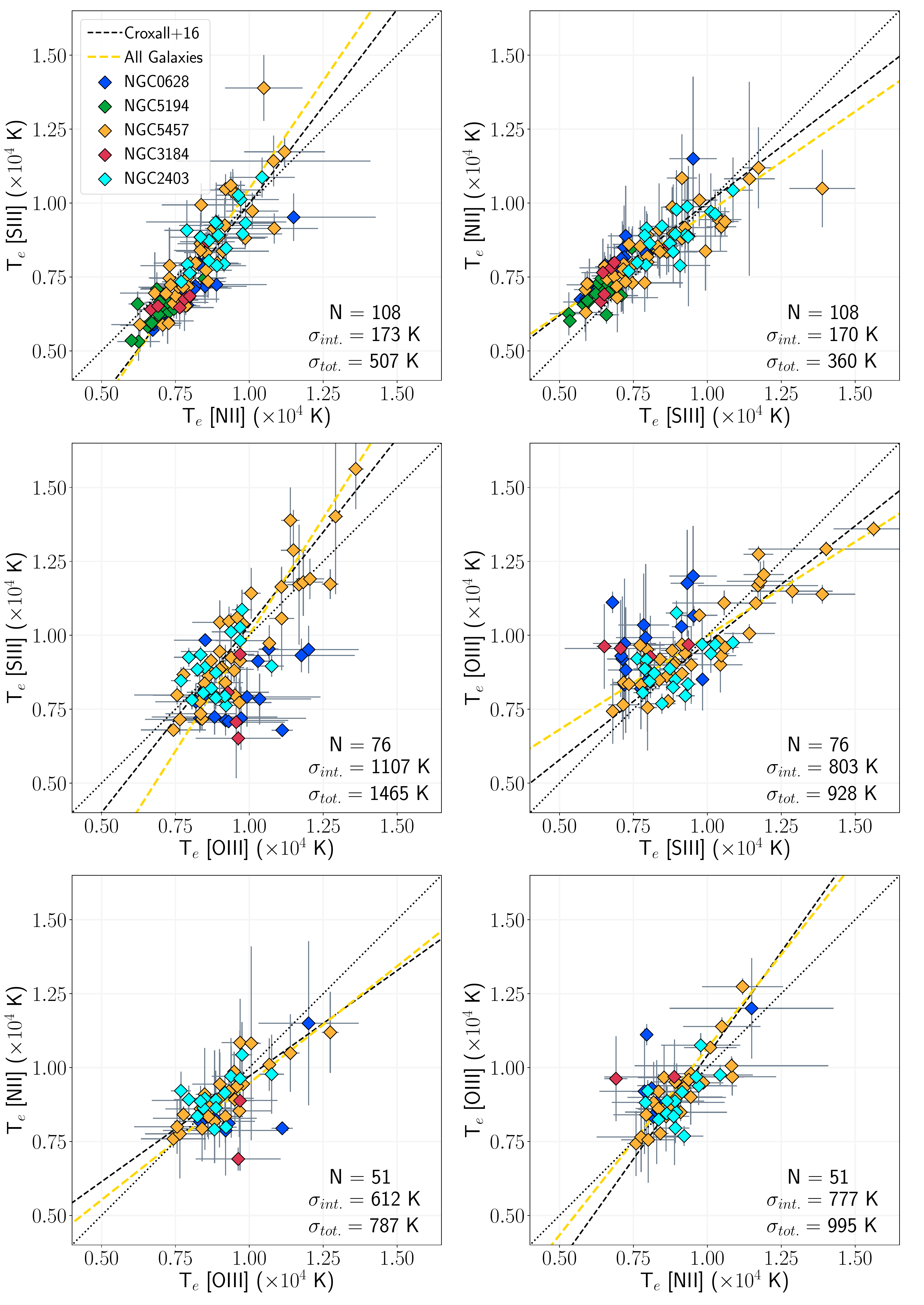}
   \caption{The T$_{e}$ data for the \oiii, \siii, and \nii\ auroral lines in each CHAOS galaxy, color coded by the host galaxy. ODR linear fits to the combined data are plotted in gold; the previous T$_e$-T$_e$ relations obtained for the NGC\,5457 H\ii\ regions \citep{croxall2016} are plotted as dashed black lines; the dotted black line represents equivalent temperatures. The rows are ordered by ionization zone relations: the top row compares temperatures in the low- and intermediate-ionization zone, the middle row compares temperatures in the intermediate- and high-ionization zone, and the bottom compares temperatures in the low- and high-ionization zones. Intrinsic and total scatters are included in the bottom-right corner of each plot.}
   \label{fig:chaosTTFig}
\end{figure*}

\subsection{CHAOS Electron Temperatures and Relations}

The combined CHAOS dataset provides an opportunity to re-examine the homogeneous T$_{e}$-T$_{e}$ relations derived from the NGC\,5457 H\ii\ regions \citep{croxall2016}. Here, we use all significant detections from \citet{berg2019} with those from NGC\,2403 to create a dataset of 213 H\ii\ regions with at least two temperature-sensitive auroral line detections. The data from \citet{berg2019} is composed of H\ii\ regions from: NGC\,628 \citep{berg2015}, NGC\,5194 \citep{croxall2015}, NGC\,5457 \citep{croxall2016}, and NGC\,3184 \citep{berg2019}. The previous T$_e$-T$_e$ relations were fit using the \textsc{Python} LINMIX package\footnote{ https://github.com/jmeyers314/linmix}. This package is the implementation of the IDL fitting program of \citet{kell2007}, and uses Bayesian statistics and MCMC techniques to fit a linear function to two variables with non-zero uncertainty while accounting for the intrinsic, random scatter in the dependent variable about the line of best-fit. However, LINMIX does not produce invertible T$_e$-T$_e$ relations due to the combination of asymmetric T$_e$ uncertainties and the package's treatment of uncertainty on the dependent and independent variables.

Here, the T$_{e}$-T$_{e}$ data from all five galaxies are fit using the \textsc{Scipy odr} package. The ODR fit to the data is assumed to be a linear relation with the data weighted by their uncertainty in both temperatures. This method provides the best-fit linear relation, but not the intrinsic dispersion about the relation. To obtain the intrinsic dispersion, we use a modified version of LINMIX. The modified version fixes the slope and intercept of the relation to the parameters of the ODR linear fit, then samples the parameter space to determine the value of $\sigma_{int}$, the intrinsic dispersion, that maximizes the likelihood function. The best-fit dispersion and its uncertainty are taken to be the median and standard deviation of the $\sigma_{int}$ distribution, respectively.

Figure \ref{fig:chaosTTFig} plots the temperatures of the auroral lines used for abundance analysis and the best-fit  T$_{e}$-T$_{e}$ relations (gold dashed lines). Each permutation of the temperature data is fit with the above method, although this technique produces invertible T$_e$-T$_e$ relations in all cases. For comparison, the black dashed lines are the best-fit T$_{e}$-T$_{e}$ relations for the CHAOS NGC\,5457 data from \citet{croxall2016}. Lines of equality are plotted as dotted black lines. The intrinsic and total dispersion \citep[the latter determined in the same manner as][]{bedr2006} in each relation are given in the bottom-right corner of each T$_{e}$-T$_{e}$ plot.

\subsubsection{Low- vs. Intermediate-Ionization \texorpdfstring{T$_e$}{Te}}

The first row of T$_{e}$-T$_{e}$ subplots in Figure \ref{fig:chaosTTFig} relates the low- and intermediate-ionization zone temperatures described by T$_{e}$\nii\ and T$_{e}$\siii\/, respectively. The addition of NGC\,2403 brings the total number of H\ii\ regions with \nii\ and \siii\ temperatures to 108. The best-fit T$_{e}$-T$_{e}$ relations are:
\begin{equation}
T_{e}\mbox{[S III]}=1.46(\pm0.07)\times\/T_{e}\mbox{[N II]} - 0.41(\pm0.05)
\end{equation}
with $\sigma_{int}=170\pm$70 K, and
\begin{equation}
T_{e}\mbox{[N II]}=0.68(\pm0.03)\times\/T_{e}\mbox{[S III]} + 0.28(\pm0.02)
\end{equation}
with $\sigma_{int}=170\pm$60 K, where the units of the two relations (and those that follow) are in 10$^{4}$ K. With the addition of the NGC\,2403 H\ii\ regions, the former relation has a slightly larger slope than the relation of \citet{croxall2016}, and the intrinsic dispersion about the fit is $\sim$100 K less than previously reported. The temperatures of NGC\,2403 further support the finding from previous CHAOS studies: a tight relation between T$_{e}$\nii\ and T$_{e}$\siii\ exists across a wide range of electron temperatures.

\subsubsection{Intermediate- vs.\ High-Ionization \texorpdfstring{T$_e$}{Te}}

The second row of panels in Figure \ref{fig:chaosTTFig} shows how the intermediate-ionization zone temperatures of T$_{e}$\siii\ are related to the high-ionization zone temperatures of T$_{e}$\oiii\/. The best-fit relations for the 76 H\ii\ regions are:
\begin{equation}
T_{e}\mbox{[S III]}=1.58(\pm0.17)\times\/T_{e}\mbox{[O III]} - 0.57(\pm0.16)
\end{equation}
with $\sigma_{int}=1110\pm$140 K, and
\begin{equation}
T_{e}\mbox{[O III]}=0.63(\pm0.07)\times\/T_{e}\mbox{[S III]} + 0.36(\pm0.06)
\end{equation}
with $\sigma_{int}=800\pm$100 K. The T$_e$\siii\/-T$_{e}$\oiii\ relation only just agrees with the previous relation of \citet{croxall2016} within uncertainty, and this relation contains the largest intrinsic dispersion about the T$_{e}$-T$_{e}$ relations of Figure \ref{fig:chaosTTFig}. Similar intrinsic dispersion in the T$_e$\siii\/-T$_{e}$\oiii\ relation is observed by \citet{croxall2016}, but the difference between this dispersion and the dispersion in the T$_e$\oiii\/-T$_{e}$\siii\ relation is not reported. The larger intrinsic dispersion in T$_e$\siii\/-T$_{e}$\oiii\ implies that the intermediate- and high-ionization zone temperatures can vary significantly due to the physical properties of an H\ii\ region.

\subsubsection{Low- vs.\ High-Ionization \texorpdfstring{T$_e$}{Te}}

The bottom row of Figure \ref{fig:chaosTTFig} plots the relations for the low- and high-ionization zone temperatures. There are fewer H\ii\ regions with concurrent \nii$\lambda$5755 and \oiii$\lambda$4363 detections than regions with concurrent \nii$\lambda$5755 and \siii$\lambda$6312 detections or \siii$\lambda$6312 and \oiii$\lambda$4363 detections. \oiii$\lambda$4363 emission corresponds to high electron energies, the product of a hard ionizing source. Depending on the ionizing source, the low-ionization zone within the H\ii\ region might be physically smaller than the high-ionization zone, making \nii$\lambda$5755 less likely to be detected. On the other hand, high-metallicity H\ii\ regions typical of spiral galaxies will have lower electron energy. Given the high excitation energy of \oiii$\lambda$4363, \nii$\lambda$5755 is more likely to be detected in these regions. Nevertheless, there are 51 H\ii\ regions with simultaneous detection, and the resulting best-fit T$_{e}$-T$_{e}$ relations are:
\begin{equation}
T_{e}\mbox{[N II]}=0.79(\pm0.14)\times\/T_{e}\mbox{[O III]} + 0.16(\pm0.13)
\end{equation}
with $\sigma_{int}=610\pm$110 K, and
\begin{equation}
T_{e}\mbox{[O III]}=1.3(\pm0.2)\times\/T_{e}\mbox{[N II]} - 0.2(\pm0.2)
\end{equation}
with $\sigma_{int}=780\pm$140 K. The first relation is consistent with the previous empirical relation for NGC\,5457, but the intrinsic dispersion is a factor of 2 larger than that of the NGC\,5457 data alone. NGC\,5457 contains the most regions with both \nii$\lambda$5755 and \oiii$\lambda$4363 detections; the inclusion of the other CHAOS galaxies adds additional scatter about the best-fit relation. The difference in the relations reveals the importance of including multiple galaxies to develop robust empirical T$_{e}$-T$_{e}$ relations.

\subsubsection{Other \texorpdfstring{T$_e$}{Te}-\texorpdfstring{T$_e$}{Te} Relations}

\begin{figure}
   \epsscale{1.2}
   \centering
   \plotone{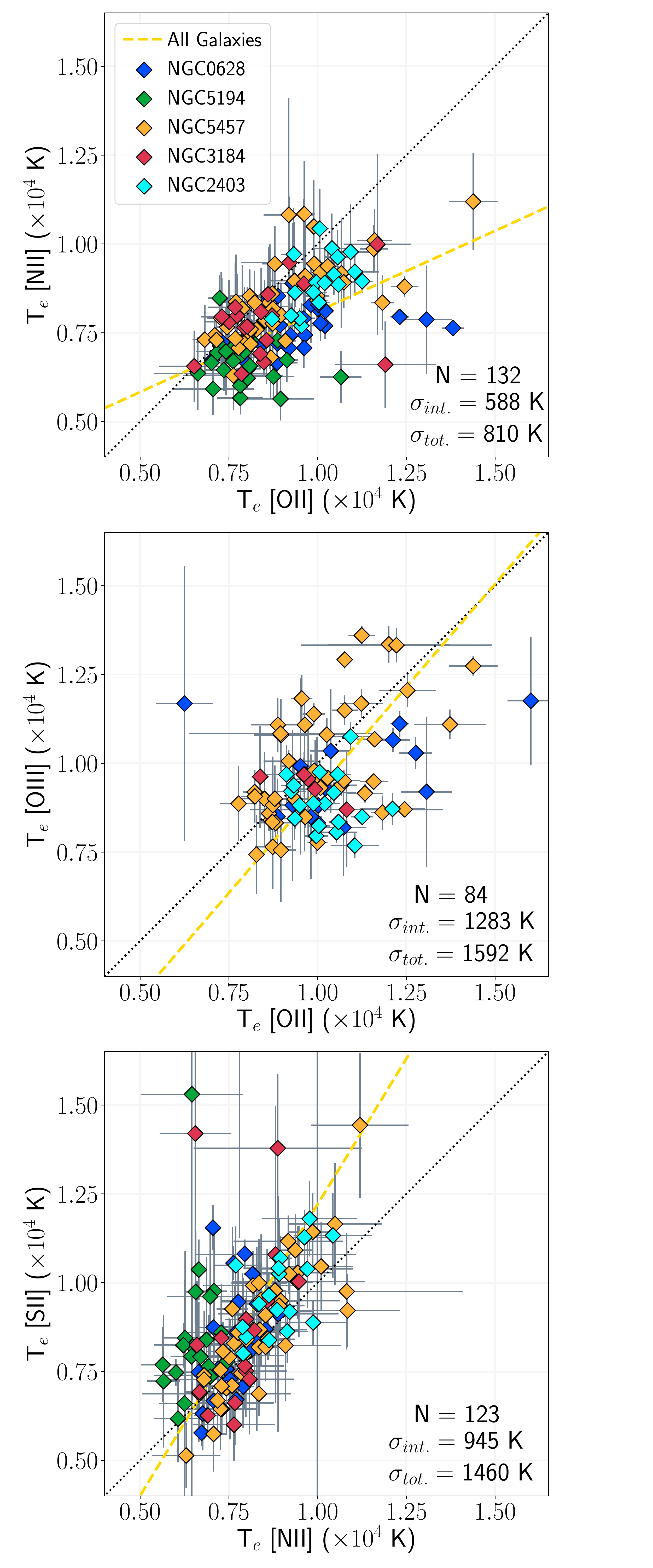}
   \caption{T$_{e}$-T$_{e}$ relations for the electron temperatures determined from the auroral lines \oii$\lambda\lambda$7320,7330 and \sii$\lambda\lambda$4069,4076. The top and bottom panels compare how these two ions compare to the low-ionization zone temperatures from \nii. The middle panel compares \oii\ temperatures to the high-ionization zone temperatures from \oiii\/.}
   \label{fig:otherTTFig}
\end{figure}

The majority of H\ii\ regions in NGC\,2403 contain \oii$\lambda\lambda$7320,7330 and \sii$\lambda\lambda$4069,4076 detections. Although these lines are presently not used to determine electron temperatures for abundance analysis, we can assess how these direct temperatures relate to those of the commonly used auroral lines described above. The top panel of Figure \ref{fig:otherTTFig} plots T$_{e}$\nii\ vs.\ T$_{e}$\oii\/, which compares the electron temperatures determined by two ions originating, primarily, in the low-ionization zone. The best-fit relation is $T_{e}\mbox{[N II]} = 0.45(\pm 0.05)\times T_{e}\mbox{[O II]} + 0.36(\pm .05)$, and the intrinsic dispersion about this relation, $\sigma_{int}=590\pm$70 K, is consistent with that found in the T$_e$\nii\ vs.\ T$_e$\oiii\/ relation. The best-fit slope is not consistent with unity, which one might expect for two ions originating in the the same ionization zone. The ODR fit prioritizes the orthogonal distance of the points with low temperature uncertainty, resulting in the relation that may not follow the majority of the data. \oii$\lambda\lambda$7320,7330 are relatively bright compared to the other auroral lines available in the optical, but the above relation reveals that it may be difficult to infer T$_e$\nii\ using \oii\ auroral line emission.

The same may be said for inferring T$_e$\oiii\ from T$_e$\oii\/, as the relation obtained in the middle panel of Figure \ref{fig:otherTTFig} has an intrinsic dispersion of $\sigma_{int}=1280\pm$140 K, larger than the dispersion observed in T$_{e}$\siii\ vs.\ T$_{e}$\oiii\/. Other abundance studies have found similar scatter about the T$_e$\oiii-T$_e$\oii\ relation \citep[see][]{kenn2003b}, and it has been shown that factors such as dielectronic recombination \citep{rubi1986,liu2001} or contamination (by airglow or telluric absorption) can affect the electron temperatures determined by \oii$\lambda\lambda$7320,7330. Dielectronic recombination biases T$_e$\nii\ and T$_e$\oii\ high, but the tight relation with \siii\ temperatures requires that the contribution of recombination to \nii$\lambda$5755 emission must be small in the majority of the H\ii\ regions observed \citep[see discussion in][]{berg2019}. A relation that is linear in electron temperature may not completely capture the trends observed in T$_{e}$\oiii\ vs.\ T$_{e}$\oii\ \citep[see][]{lope2012,nich2014,yate2020}. For now, we simply note that the \oii\ temperatures measured in NGC\,2403 are consistent with the trends found in the other CHAOS studies. 

T$_{e}$\sii\ vs.\ T$_{e}$\nii\/, which also examines the relation of two ions in the low-ionization zone, is plotted in the bottom of Figure \ref{fig:otherTTFig}. This relation, $T_{e}\mbox{[S II]} = 1.64(\pm 0.17)\times T_{e}\mbox{[N II]} - 0.42(\pm .13)$, is also not consistent with a slope of unity. The general trend is an offset toward higher T$_{e}$\sii\ at fixed T$_{e}$\nii\/, and the intrinsic dispersion is large: $\sigma_{int}=950\pm$120 K. This scatter is possibly related to the differences in the ionization energies of the two ions. \sii\ emission can originate in a photodissociation region (PDR), a region of primarily neutral gas outside the H\ii\ region. The temperature within a PDR is not characteristic of the electron temperature within the H\ii\ region, so a comparison between the two may result in the increased scatter in T$_e$-T$_e$ space observed in the bottom panel of Figure \ref{fig:otherTTFig}.

More direct electron temperature data are needed to fully explore the empirical T$_{e}$-T$_{e}$ relations. For example, the tight relation between T$_{e}$\nii\/-T$_{e}$\siii\ may not hold at larger temperatures where there is currently a lack of direct T$_{e}$\nii\ and T$_{e}$\siii\ data. For intermediate- and high-ionization zone relations, \citet{berg2019} note that the offsets in T$_{e}$-T$_{e}$ space from the best-fit relation could be dependent on an H\ii\ region’s average ionization, characterized by the parameter $O_{32} = \frac{I(5007)}{I(3727)}$. An analysis of how these offsets are dependent on second parameters will be conducted once more electron temperature data are added to the CHAOS sample. Finally, the trends found in \oii\ and \sii\ temperatures can be explored with the entire CHAOS sample, with the hope that future abundance studies will make use of these temperatures.

\subsection{Application of CHAOS \texorpdfstring{T$_e$}{Te}-\texorpdfstring{T$_e$}{Te} Relations}

As described above, applying a linear T$_{e}$-T$_{e}$ relation to obtain the electron temperature in a different ionization zone may not account for the unique physical properties within a given H\ii\ region. If the intrinsic dispersion in T$_{e}$-T$_{e}$ space is representative of how differences in physical properties of H\ii\ regions affect the measured electron temperatures in these zones, then one must account for this dispersion in order to appropriately infer the temperature in another zone. Not accounting for this intrinsic dispersion is equivalent to neglecting the shortcomings of a linear relation in fitting the data and to assuming that specific H\ii\ region properties have little effect on the dispersion of measured electron temperatures.

The uncertainties on the electron temperature inferred from a T$_e$-T$_e$ relation present an opportunity to incorporate the intrinsic dispersion about the T$_e$-T$_e$ relation. Previous abundance studies have obtained uncertainties on inferred temperatures from T$_{e}$-T$_{e}$ relations in a number of different ways. For instance, \citet{skil2003} used standard propagation of errors to obtain the uncertainties on the inferred temperatures when applying the photoionization temperature relations of \citet{page1992}, and set a lower limit on the inferred T$_e$ uncertainty of 500 K. The lower limit is applied to avoid the small uncertainties on inferred temperatures that would result from propagating errors through a linear T$_e$-T$_e$ relation that does not account for all the physical properties of an H\ii\ region. \citet{kenn2003b} used either the uncertainty on the measured temperature as the uncertainty on the inferred temperature when applying the \citet{garn1992} T$_e$-T$_e$ relations, or augment this uncertainty by 500 K added in quadrature. \citet{roso2008} used the photoionization T$_{e}$- T$_{e}$ relations of \citet{camp1986} to obtain T$_e$\oii\ for all H\ii\ regions they observe in M33. The error in their inferred T$_e$\oii\ values is assumed to be 300 K, citing the magnitude of the uncertainties in \citet{kenn2003b} as justification for this choice. Finally, \citet{este2020} used the empirical, linear T$_{e}$-T$_{e}$ relations from \citet{este2009} when necessary, and standard propagation of errors is applied, without a lower limit, to obtain the uncertainties on the inferred temperatures.

With the amount of electron temperature data amassed by CHAOS, we can now update the method we apply to obtain uncertainties on inferred electron temperatures. This method attempts to account for both the uncertainty in the measured temperature being used to infer a temperature in a different ionization zone, and the uncertainty in applying a linear T$_e$-T$_e$ relation to account for the potential range of physical conditions within an H\ii\ region. For an inferred electron temperature $T_{e,Y}$ determined by the direct temperature $T_{e,X}$ and the T$_{e}$-T$_{e}$ relation $T_{e,Y} = m\times\/T_{e,X} + b$, the uncertainty on the inferred temperature, $\delta T_{e,Y}$, is now determined by:
\begin{equation} \label{eq:dT_eq}
\delta T_{e,Y} = \sqrt{(m\times\delta T_{e,X})^{2} + (\sigma_{i,Y})^{2}},
\end{equation}
where $\sigma_{i,Y}$ is the intrinsic dispersion in $T_{e,Y}$ about the relation. With this equation, a lower bound is imposed on the inferred temperature uncertainty equivalent to the intrinsic dispersion about the best-fit relation. In this way, applying one of the above T$_{e}$-T$_{e}$ relations always results in a higher fractional uncertainty on the inferred temperature, but the result accounts for some of the unknowns that are not fit when applying a linear T$_{e}$-T$_{e}$ relation. In other words, the uncertainty on the inferred temperature is now more likely to capture the true electron temperature within the ionization zone. The best-fit variables and intrinsic dispersion of each relation will change as more electron temperature data are acquired; if the intrinsic dispersion becomes smaller, then so too will the lower bound on the inferred temperature uncertainties. Moving forward, we use Equation \ref{eq:dT_eq} to determine the uncertainty in the inferred electron temperatures, and this approach is recommended for all studies using linear, empirical T$_{e}$-T$_{e}$ relations.

While the intrinsic dispersion about a T$_{e}$-T$_{e}$ relation is readily obtained for empirical relations, an estimate on the scatter when using photoionization model T$_e$-T$_e$ relations can be obtained by varying model inputs and finding a range of possible best-fit parameters. More recent T$_e$-T$_e$ relations have adopted non-linear forms \citep{lope2012,arel2020}, while others have included a dependence on the H\ii\ region's metallicity \citep{nich2014,yate2020}. An exploration of how the intrinsic scatter can be used in these relations is beyond the scope of this paper, but we stress that inferred electron temperatures that account for the two main sources of uncertainty discussed above are critical for proper abundance analysis.

In fitting the different temperature permutations with a linear relation, we have assumed that each ion temperature is a smooth function of the electron temperature in a different ionization zone. As mentioned earlier, ionization zone temperatures within an H\ii\ region are dependent on similar parameters, such as degree of ionization, metallicity, etc. Additionally, the errors on the data strongly affect the ODR linear fits, which can bias the line of best fit towards temperatures from well-measured lines in extremely bright H\ii\ regions. In future works, we will consider additional parameters and explore possible non-linear T$_e$-T$_e$ relations to better fit the observed scatter in electron temperatures. 

%%%%%%%%%%%%%%%%%%%%%%%%%%%%%
\section{Abundance Determinations in NGC 2403}

To determine gas-phase abundances, we assume a five-level atom model \citep{dero1987} with the updated atomic data used in \citet{berg2015}. This model is used with the electron temperature of a given ionization zone in \textsc{PyNeb}’s getIonAbundance function, assuming an electron density of 10$^{2}$ cm$^{-3}$, to obtain the abundance of an ion in that zone. The fractional uncertainty in the line intensity ratio and in the emissivity (as a function of temperature) are added in quadrature to obtain the ionic abundance uncertainty. This process is applied to the uncertainty on relative abundances of ions within the same ionization zone (e.g., Ne$^{2+}$/O$^{2+}$). For these cases, the fractional uncertainty on the net emissivity is usually small relative to the intensity uncertainty because of the similar temperature dependencies of the emissivities.

Previously, \citet{berg2019} used ionization-based temperature prioritizations to determine the electron temperature in the three ionization zones. This method attempts to use, when possible, a direct temperature measurement from the dominant ionization zone within an H\ii\ region \citep[see Figure 5 in][reproduced in Appendix \ref{sec:A2} as Figure \ref{fig:temp_priors}]{berg2019}. If a temperature outside the dominant ionization zone is used with a T$_e$-T$_e$ relation, then the resulting fractional uncertainty on the ionization zone temperature will be large due to the addition of the intrinsic dispersion about the T$_e$-T$_e$ relation. This increased temperature, and abundance, uncertainty is sometimes not reflective of the quality of the spectra, particularly spectra with multiple auroral line detections. For instance, an \oiii$\lambda$4363 detection at high S/N might more reliably determine the true electron temperature in the high-ionization zone than a low S/N \siii$\lambda$6312 detection with a T$_e$-T$_e$ relation.

An alternative approach is to use all electron temperature measurements as independent methods of calculating the electron temperature in a given ionization zone. This is justified because all ionization zones within an H\ii\ region are assumed to have the same metallicity, and metallicity is the dominant parameter in determining the electron temperature. Thus, electron temperatures in different ionization zones are strongly correlated. When all three commonly used auroral lines are measured, the direct electron temperature from the dominant ion and the two inferred temperatures from the T$_e$-T$_e$ relations are combined in a weighted average to determine the ionization zone temperature. The uncertainty on an inferred electron temperature when using a T$_e$-T$_e$ relation will typically weight these temperatures lower than a direct temperature measurement from the dominant ion. The exception is when the dominant ion is measured at low S/N, which results in an ionization zone temperature that is closer to the average of the measured and inferred temperatures. In this way, a single, poorly constrained electron temperature will not bias the ionization zone temperature used in abundance determination.

The uncertainty in the ionization zone temperature is taken to be the uncertainty of the weighted average, which is smaller than the uncertainty on the most well-measured temperature used in calculating the weighted average. The abundance analysis has been repeated by adopting the lowest temperature uncertainty of the values used to calculate the weighted average, and all results are consistent within uncertainty. Adopting a weighted average ionization zone temperature removes the prioritization based on the average ionization of an H\ii\ region in favor of utilizing all available temperature data in the CHAOS sample. A comparison of the weighted average and ionization-based temperature prioritizations is given in Appendix \ref{sec:A2}, and it is found that the two prioritization methods produce consistent results. As discussed in \S3.2.4, the \oii\ and \sii\ auroral lines are not used for abundance determination due to the dispersion observed in the temperatures. Only the H\ii\ regions with at least one of the auroral lines from \oiii\/, \nii\/, or \siii\ are used for abundance determination, which brings the number of H\ii\ regions in NGC\,2403 that are used for abundance analysis to 28. Table \ref{t:abundances} provides the adopted temperature in each ionization zone, the ionic abundances, and Ionization Correction Factors (ICFs) used for these 28 regions. 

\subsection{Oxygen Abundances}

The dominant ionization states of oxygen in an H\ii\ region are O$^{+}$ in the intermediate- and low-ionization zones and O$^{2+}$ in the high-ionization zone. The ionization energy of O$^{0}$ is close to that of neutral hydrogen, but the ratio of [\ion{O}{1}]$\lambda$6300 to H$\beta$ can assess the amount of O$^{0}$ in each region. However, the observed neutral oxygen emission may come from a PDR, so including the neutral oxygen may not appropriately estimate the amount of oxygen contained within the H\ii\ region.

For the typical temperature of an O or B type star, the number of photons able to triply ionize oxygen make up a small fraction of the total photons produced. Photons that can triply ionize oxygen can doubly ionize helium, so we would expect emission from the \ion{He}{2} recombination lines if we observe an extremely ionized H\ii\ region. For the one H\ii\ region where narrow \ion{He}{2} $\lambda$4686 emission is detected (NGC2403+160$-$251), a possible correction to account for the presence of O$^{3+}$/H$^+$ could be justified. However, the size of this correction is smaller than the uncertainty on O/H, so no correction is applied. Additionally, we do not correct for depletion of oxygen onto dust grains, which is on the order of 0.1 dex in the most metal-rich H\ii\ regions \citep{peim2010,pena2012}. Therefore, it is assumed that all of the oxygen in an H\ii\ region is present in either O$^{+}$ or O$^{2+}$ such that O/H = (O$^{+}$+O$^{2+}$)/H$^{+}$.

\subsection{Nitrogen Abundances}

Other common emission lines from N, Ne, S, and Ar are observed in a typical CHAOS spectrum, but these elements have unobserved ionic species in the optical. Ionization Correction Factors account for the unobserved ionic species of an element, J, by weighting the observed ionic species with a function, ICF(J), or: $\frac{\text{J}}{\text{H}} =\text{ICF(J)}\times\frac{\text{J}^{+\text{i}}}{\text{H}^{+}}$. A common example is nitrogen with the observable lines of N$^{+}$ but with no emission lines of N$^{2+}$ or N$^{3+}$ in our wavelength range. The ionization energy of O$^{+}$ fully spans that of N$^{+}$, thus it is often assumed that N/O $\approx$ N$^{+}$/O$^{+}$. This approximation is useful because the ratio primarily relates two lines originating in the low-ionization zone, but O$^{+}$ also slightly overlaps with the intermediate-ionization zone. This ICF is found to be good to within 10\% of the actual N/O abundance, with the departures coming primarily from low-metallicity (12+log(O/H) $<$ 8.1) systems \citep{nava2006}. None of the H\ii\ regions in NGC\,2403 have oxygen abundances less than 8.1, so it is assumed that ICF(N) = O/O$^+$ will accurately describe the data. We determine the N$^+$/O$^+$ relative abundance and uncertainty using the emissivity ratio j$_{\textrm{[N II]}6584}$/j$_{\textrm{[O II]}3727}$ directly with the low-ionization zone temperature and the observed intensity of \nii$\lambda$6584 and \oii$\lambda$3727.

\subsection{\texorpdfstring{$\alpha$}{a} Elements: ICFs and Abundances}

\subsubsection{Neon}

\begin{figure}
   \epsscale{1.17}
   \centering
   \plotone{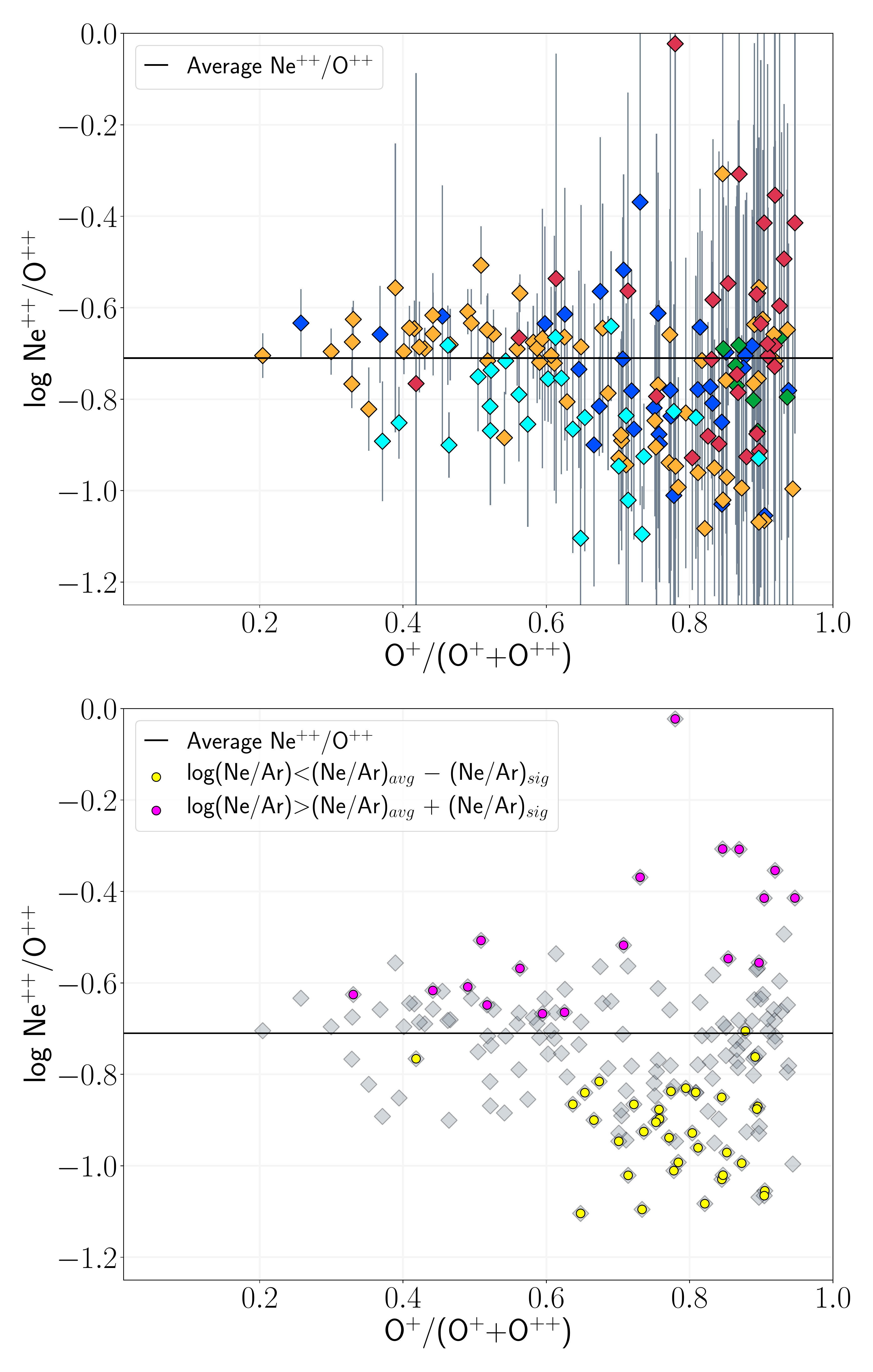}
   \caption{\textit{Top Panel:} log(Ne$^{2+}$/O$^{2+}$) data from all five CHAOS galaxies plotted against each region's O$^{+}$/O. Uncertainty on O$^{+}$/O is not plotted for clarity. The weighted average of the data is plotted as a solid black line. Data is color coded by galaxy (same as Figure \ref{fig:chaosTTFig}). \textit{Bottom Panel:} The same data is plotted, except the colors and errors are now removed. The yellow points indicate regions with log(Ne/Ar) less than 1$\sigma$ below the average log(Ne/Ar) of all regions. Purple points indicate regions with log(Ne/Ar) greater than 1$\sigma$ above the average. The dispersion about the average Ne$^{2+}$/O$^{2+}$ grows at lower ionization and is dominated by the points with enhanced or reduced Ne/Ar.}
   \label{fig:neICF}
\end{figure}

The singly- and doubly-ionized states of neon are present in a typical H\ii\ region, but only one line, [\ion{Ne}{3}]$\lambda$3868, is commonly, and easily, observed in CHAOS spectra\footnote{The reduction pipeline also fits [\ion{Ne}{3}]$\lambda$3967, but this line is blended with \ion{He}{1} $\lambda$3964 and H7 $\lambda$3970. As such, it is not used for Ne abundance determinations.}. [\ion{Ne}{3}] emission comes from the high-ionization zone, so it is common to employ the ICF of \citet{peim1969}: ICF(Ne) = O/O$^{2+}$, such that Ne/O = Ne$^{2+}$/O$^{2+}$ \citep[see also][]{croc2006}. However, the reliability of this ICF at low ionization has come into question. This can be seen in the top panel of Figure \ref{fig:neICF}, where the scatter about the best-fit average (black line) of Ne$^{2+}$/O$^{2+}$ grows substantially at intermediate ionization (O$^{+}$/O $>$ $0.5$), with much of the scatter coming from regions with low Ne$^{2+}$/O$^{2+}$ at fixed O$^{+}$/O. This trend has been observed in previous CHAOS studies \citep{croxall2016,berg2019} and planetary nebula studies \citep[for example,][]{torr1977}.

Motivated by the findings of \citet{garc2013}, we examine Ne/Ar in each region and find that many of the regions with low Ne$^{2+}$/O$^{2+}$ have low Ne/Ar. For regions with O$^{+}$/O $>$ $0.5$, \citet{berg2019} corrected the Ne/O abundance by subtracting the difference between log(Ne/Ar) and log(Ne/Ar)$_{avg}$ from log(Ne/O). For regions with low Ne/Ar relative to the average Ne/Ar of all regions, this preliminary correction offsets the low neon abundances and results in larger log(Ne/O) at low ionizations. Alternatively, the Ne/O abundances can be lowered for regions with high Ne/Ar (or low Ar/O) abundances.

In the lower panel of Figure \ref{fig:neICF}, the regions with log(Ne/Ar) $<$ log(Ne/Ar)$_{avg}$ $-$  log(Ne/Ar)$_{sig}$ are noted in yellow, where log(Ne/Ar)$_{sig}$ is the 1$\sigma$ error on log(Ne/Ar)$_{avg}$. These are the regions that would benefit most from the correction in \citet{berg2019}, resulting in larger Ne$^{2+}$/O$^{2+}$ values and lower scatter about the best-fit Ne$^{2+}$/O$^{2+}$ average. The regions with log(Ne/Ar) $>$ log(Ne/Ar)$_{avg}$ +  log(Ne/Ar)$_{sig}$ are noted in purple. These regions could benefit from the correction, but some regions close to the line of best fit have low argon abundance (see Figure \ref{fig:arICF}) resulting in relatively large offsets from the average Ne/Ar value. This means that these regions could have their Ne$^{2+}$/O$^{2+}$ ratios over-subtracted, resulting in increased scatter \citep[this has also been noted in][]{berg2019}. We simply adopt ICF(Ne) = O/O$^{2+}$ for this study. We use the relative emissivities of [\ion{Ne}{3}]$\lambda$3868 and \oiii$\lambda$5007, the intensity of these lines, and the high-ionization zone temperature to obtain the Ne$^{2+}$/O$^{2+}$ relative abundance.

\subsubsection{Sulfur}

\begin{figure}
   \epsscale{1.17}
   \centering
   \plotone{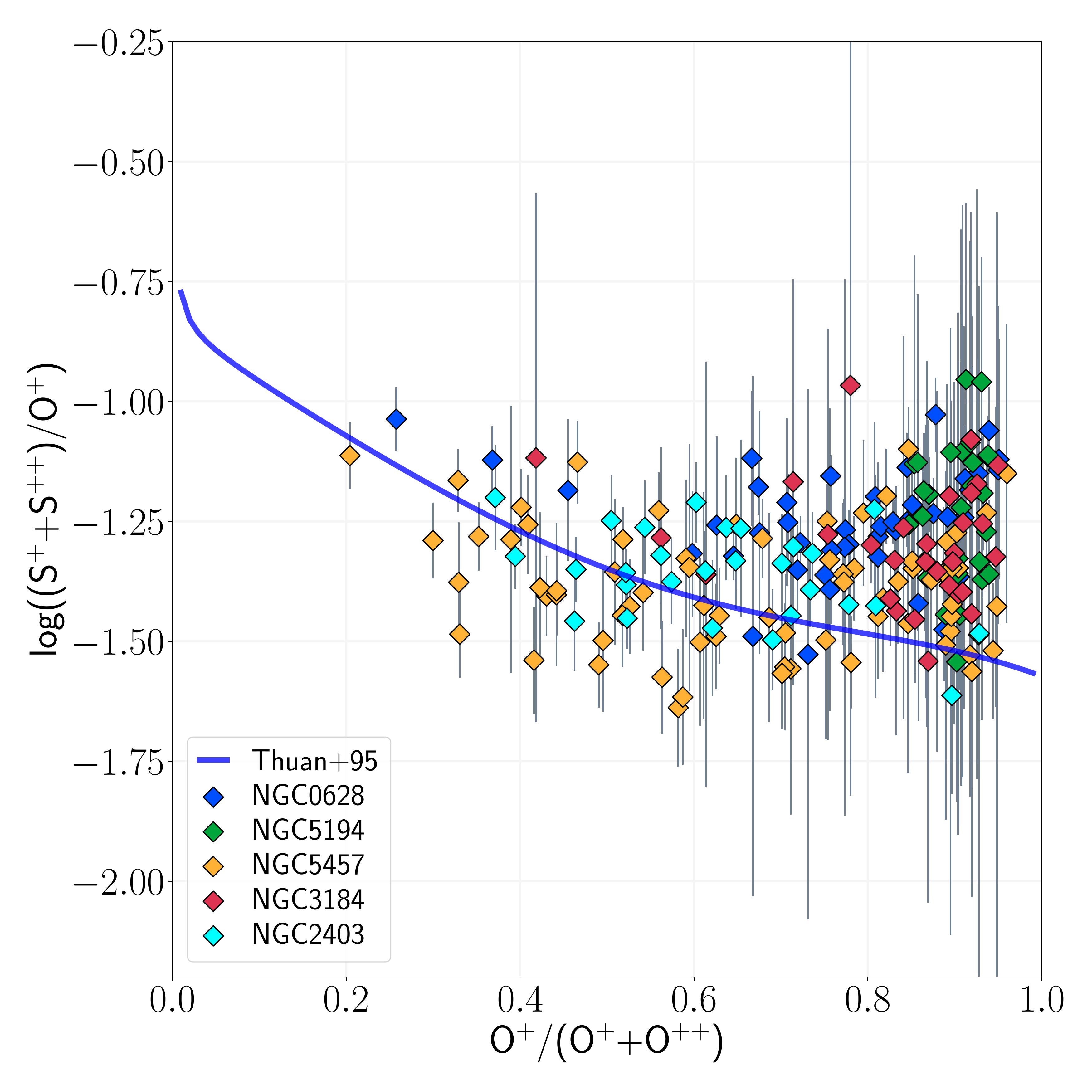}
   \caption{The log((S$^{+}$ + S$^{2+}$)/O$^{+}$) data from all five CHAOS galaxies plotted against each region's O$^{+}$/O. Uncertainty on O$^{+}$/O is not plotted for clarity. The \citet{thua1995} photoionization ICF is plotted as a solid blue line. While the photoionization ICF appears to under-predict the (S$^{+}$ + S$^{2+}$)/O$^{+}$ at low-ionization, it fits the general trend of the regions with O$^+$/O $<$ 0.6.}
   \label{fig:sICF}
\end{figure}

The emission from S$^{2+}$ characterizes the intermediate-ionization zone, but S$^{3+}$ is expected to be present in an H\ii\ region given that the ionization energy of S$^{2+}$ is $34.79$ eV, about the same as that of O$^{+}$. However, there are no S$^{3+}$ emission lines in our wavelength range, requiring an ICF to account for the missing ions in this ionization state. The ionization energies of S$^{0}$ and S$^{2+}$ ($10.36$ and $34.79$ eV) cover an energy range that is nearly coincident with that of O$^{0}$ and O$^{+}$ ($13.62$ and $35.12$ eV). Previous CHAOS studies \citep{croxall2016,berg2019} have examined the approximation S/O = (S$^{+}$ + S$^{2+}$)/O$^{+}$ \citep[see][]{peim1969} such that the ICF for S is ICF(S) = O/O$^{+}$ = (O$^{+}$ + O$^{2+}$)/O$^{+}$. This approximation is particularly useful when the ionization of an H\ii\ region is such that the O$^{+}$ zone is more dominant than the O$^{2+}$ zone. In such an H\ii\ region, the ICF is accounting for a smaller fraction of unobserved S$^{3+}$ and is using detections from a proportionally larger volume of the H\ii\ region to infer the relative sulfur abundance. For higher ionization H\ii\ regions, this ICF is not adequate to describe the amount of S$^{3+}$ present. 

Figure \ref{fig:sICF} plots log((S$^{+}$ + S$^{2+}$)/O$^{+}$) vs.\ O$^{+}$/O for the H\ii\ regions in the present CHAOS sample. The sulfur ICF of \citet{thua1995} is plotted as a solid blue line. This ICF is generated from the photoionization models of \citet{stas1990} and is a function of O$^{+}$/O. The ICF tends to underpredict the (S$^{+}$ + S$^{2+}$)/O$^{+}$ observed in the CHAOS H\ii\ regions with O$^{+}$/O $>$ 0.6, which are the regions where the ICF(S) = O/O$^{+}$ is believed to be the most reliable. Instead, we find general agreement between the ICF of \citet{thua1995} and the data at O$^{+}$/O $<$ $0.6$, although there are fewer highly-ionized H\ii\ regions in the sample and there is an appreciable amount of scatter. Given this general agreement, we adopt the ICF(S) = O/O$^{+}$ when O$^{+}$/O $>$ $0.6$ and the ICF from \citet{thua1995} when O$^{+}$/O $<$ $0.6$, similar to the technique previously applied in other CHAOS galaxies \citep[see][]{croxall2016,berg2019}. The percent uncertainty when applying the \citet{thua1995} ICF is assumed to be 10\%. The energy required to ionize S$^{0}$ ($10.36$ eV) is such that some S$^{+}$ is contained within a PDR. This portion should be discounted when determining the true sulfur abundance within the H\ii\ region, but it is assumed that this makes up a small fraction of the total S$^{+}$ abundance.

\subsubsection{Argon}

\begin{figure}
   \epsscale{1.17}
   \centering
   \plotone{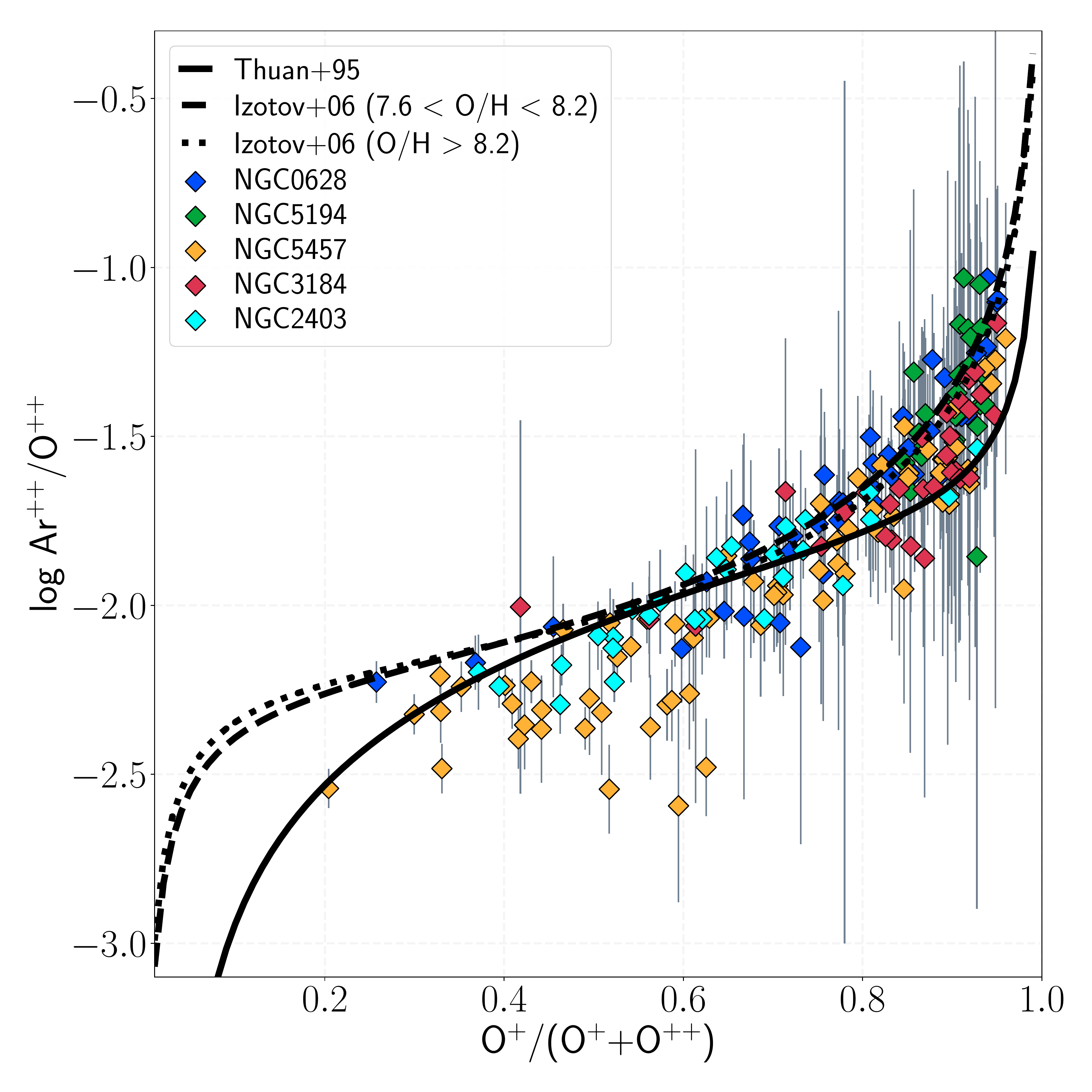}
   \caption{The log(Ar$^{2+}$/O$^{2+}$) data from all five CHAOS galaxies plotted against each region's O$^{+}$/O. Uncertainty on O$^{+}$/O is not plotted for clarity. With the updated temperatures and abundances, the shape of the CHAOS data is better fit by the metallicity-dependent ICF of \citep{izot2006} (dashed and dotted lines) rather than the ICF of \citet{thua1995} (solid line).}
   \label{fig:arICF}
\end{figure}

Multiple ionization states of Ar are expected to be present in a typical H\ii\ region: Ar$^{+}$ is present in the low-ionization zone, Ar$^{2+}$ in the intermediate- and high-ionization zones, and Ar$^{3+}$ can be found in the high-ionization zone. For CHAOS, the only easily observed emission lines in these ionization states are [Ar \,{\sc iii]}$\lambda$7135 and the occasional [\ion{Ar}{4}]$\lambda$4740 in the most highly ionized H\ii\ regions. The overlap of Ar$^{2+}$ with the intermediate-ionization zone has motivated previous studies to use Ar$^{2+}$/S$^{2+}$ relative abundances and ICFs. Both \citet{kenn2003b} and \citet{croxall2016} found that the Ar$^{2+}$/S$^{2+}$ ratio remains relatively constant over a range of O$^{+}$/O values, with the latter adopting a linearly-decreasing ICF to account for the lowest ionization H\ii\ regions of the CHAOS sample. After updating the ionic abundance data using ionization-based temperature prioritizations, \citet{berg2019} determined that the photoionization model ICF from \citet{thua1995} fit the CHAOS data over a large range of ionization.

Figure \ref{fig:arICF} plots the Ar$^{2+}$/O$^{2+}$ data vs.\ O$^{+}$/O from the CHAOS sample. With the updated electron temperatures and the use of a weighted average temperature in each ionization zone, the agreement between the CHAOS data and the expected Ar$^{2+}$/O$^{2+}$ trend from the \citet{thua1995} ICF (solid line) is less obvious, particularly at O$^+$/O $>$ 0.8. Applying this ICF may lead to unphysical trends in Ar/O, requiring a new argon ICF. Plotted as dashed and dotted lines are the intermediate- and high-metallicity argon ICFs of \citet{izot2006}, respectively. The shape of these ICFs match the observed trends in the CHAOS Ar$^{2+}$/O$^{2+}$ data over nearly the entire range of H\ii\ region ionization. We update our Ar$^{2+}$ ICF to be that of \citet{izot2006}. The intermediate-metallicity ICF is applied when an H\ii\ region has 12+log(O/H) $<$ 7.6, the high-metallicity ICF is applied at 12+log(O/H) $>$ 8.2, and a linear interpolation of the two for 7.6 $<$ 12+log(O/H) $<$ 8.2. As with the sulfur ICF, there is an attributed 10\% uncertainty for applying the \citet{izot2006} ICF.

%%%%%%%%%%%%%%%%%%%%%%%%%%%%%
\section{Direct Abundance Gradients}

The above methods are used to calculate the abundances of all CHAOS galaxies using all regions from \citet{berg2019} with at least one of the auroral lines from \oiii\/, \nii\/, or \siii\ detected. The following sections discuss the resulting O/H and N/O abundance gradients

\subsection{Oxygen Abundance Gradient in NGC 2403}

\begin{figure}
   \epsscale{1.17}
   \centering
   \plotone{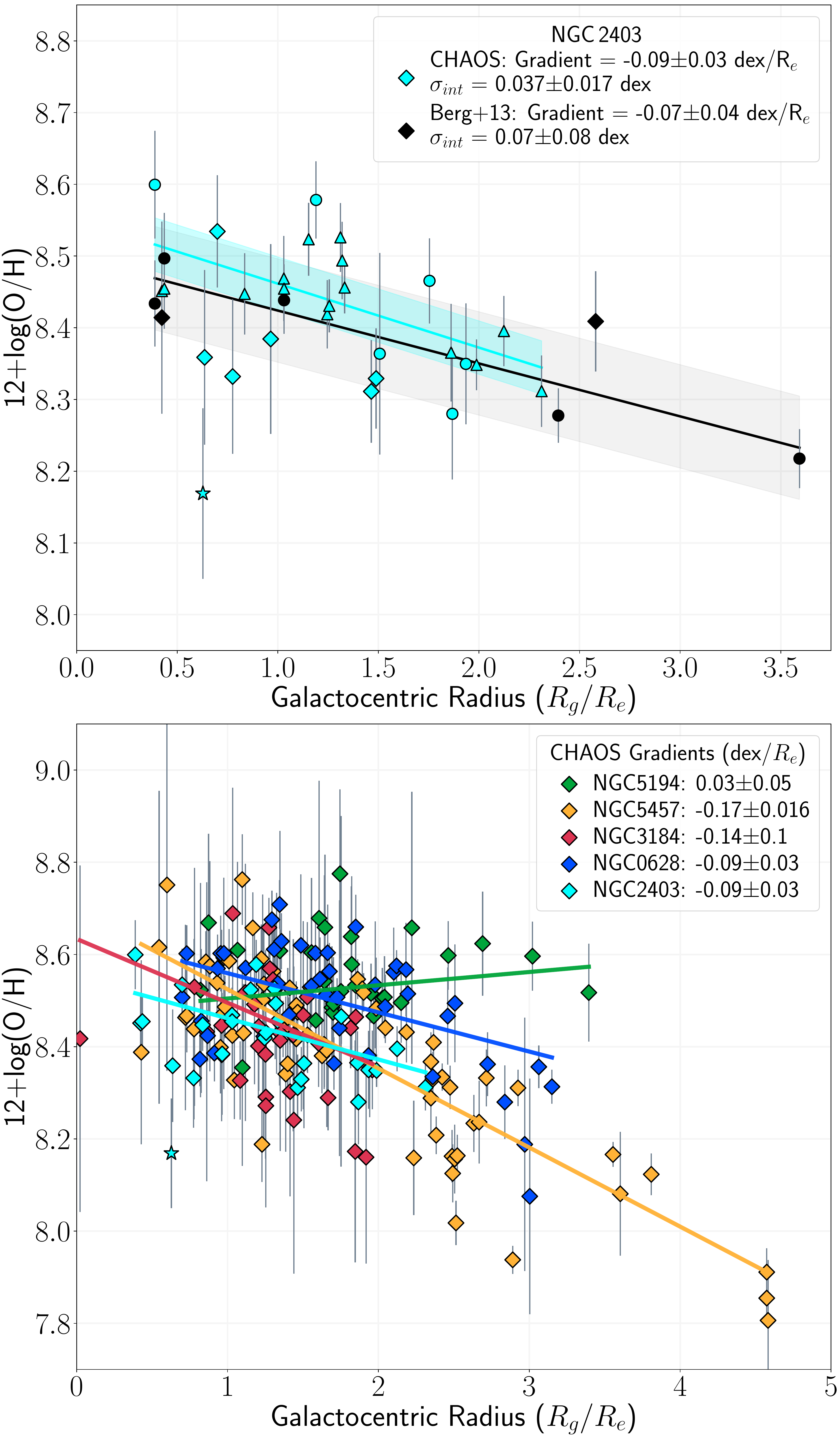}
   \caption{\textit{Top Panel:} CHAOS (cyan) and the recalculated \citet{berg2013}  (black) oxygen abundances in NGC\,2403 plotted vs.\ R$_g$/R$_e$. The gradients in each galaxy are plotted as solid lines and are provided in the legend. The intrinsic dispersion about each gradient, represented as the shaded region around each gradient, are also reported in the legend. The number of direct electron temperatures used in the weighted average temperature in each ionization zone are represented by the different shapes: Diamonds = 1 Direct Temperature; Circles = 2; Triangles = 3. \textit{Bottom Panel:} Same as top panel but for the oxygen abundances of all CHAOS galaxies and the regions are no longer distinguished by the number of direct T$_e$ used in the average. The shaded portions around each gradient are removed for clarity, and the galaxies listed in the legend are ordered by decreasing stellar mass.}
   \label{fig:ohAbun}
\end{figure}

The top panel of Figure \ref{fig:ohAbun} plots, in cyan, the oxygen abundances measured in NGC\,2403. The abundances are plotted against the galactocentric radius of the H\ii\ regions normalized to the effective, or half-light, radius (R$_{e}$) of NGC\,2403. The effective radius of NGC\,2403 is measured from the Z0MGS WISE 1 maps using the same technique as applied to the previous four CHAOS galaxies \citep[see discussion in Appendix C of][]{berg2019}. For each region, the number of direct electron temperatures used in the weighted average ionization zone temperatures is designated by the shape of the point used: Diamonds = 1 temperature, Circles = 2, and Triangles = 3.

LINMIX is applied to find the best-fit gradients, assuming that the errors on the distances are 0.05$\times$R$_{e}$ for each H\ii\ region. Additionally, the abundance gradients are fit with the ODR and modified LINMIX combination described in \S3.2. The abundance gradients and intrinsic dispersions determined by the LINMIX-only fits are consistent with the ODR-LINMIX combination fits, so we adopt the former. The gradient determined for the regions of NGC\,2403 reported here is:
\begin{equation}
\mbox{12+log(O/H)}=8.55(\pm0.04) - 0.09(\pm0.03)R_{g}/R_{e}.
\label{eq:ohabun}
\end{equation}
Using the position of each H\ii\ region in kpc, the gradient in NGC\,2403 is
\begin{equation}
\mbox{12+log(O/H)}=8.55(\pm0.04) - 0.032(\pm0.010)R_{g}/kpc.
\end{equation}
The intrinsic dispersion about these fits is $\sigma_{int}=0.037\pm0.017$ dex; the shaded region about the gradient in Figure \ref{fig:ohAbun} represents the intrinsic dispersion. The oxygen abundance gradient is determined from 27 H\ii\ regions; one H\ii\ region, NGC2403+88$-$18 (the cyan star in Figure \ref{fig:ohAbun}), is not fit in the reported oxygen abundance gradient due to the presence of unrecognizable and extreme emission features.

\citet{garn1997} and \citet{berg2013} have previously observed NGC\,2403 and conducted direct abundance studies on a number of H\ii\ regions within the galaxy. The former used the Imaging Photon Counting System (IPCS) at La Palma Observatory to measure the auroral lines of \oiii, \siii, and \oii\ in 12 H\ii\ regions in NGC\,2403. The latter used the Blue Channel Spectrograph on the MMT to observe seven bright H\ii\ regions in NGC\,2403, allowing for direct T$_e$\oiii\ and T$_e$\nii\ determination. These two studies measured the slope of the oxygen abundance gradient in NGC\,2403 as $-0.102\pm0.009$ dex/kpc \citep{garn1997} and $-0.027\pm0.008$ dex/kpc \citep{berg2013}. Additionally, \citet{berg2013} reported an intrinsic scatter about the oxygen abundance gradient of 0.02 dex. We observe some of the same H\ii\ regions as \citet{garn1997}, but the IPCS has non-linear counting effects and complicated statistical errors at all count rates \citep{jenk1987}. As such, we only perform a complete comparison to the direct abundances of \citet{berg2013}.

\citet{berg2013} targeted seven bright H\ii\ regions, four of which overlap with our observations. Multiple factors could result in differences between our findings and the results reported in \citet{berg2013}: our T$_{e}$-T$_{e}$ relations and uncertainties are different from those of \citet{garn1992} (see \S3); we determine the electron temperature in an ionization zone using the weighted average of all temperature data; the atomic data are updated \citep[see][]{berg2015}; and we observe roughly four times as many regions with temperature-sensitive auroral lines. To eliminate as many systematic differences as possible, we use the reported line intensities from the H\ii\ regions in \citet{berg2013} and recalculate the temperatures and abundances following the methods described in \S3 and \S4. This includes the use of \textsc{PyNeb} to determine T$_e$\nii\ and T$_e$\oiii\ from their observations (T$_e$\siii\ is unobtainable due to the lack of wavelength coverage for the \siii\ nebular lines), the T$_e$-T$_e$ relations and new application method described in \S3, weighted average temperatures in each ionization zone, and updated atomic data and ICFs for the elemental abundances.

The top panel of Figure \ref{fig:ohAbun} plots, in black, the updated direct abundances of NGC\,2403 acquired from the line intensities of \citet{berg2013}\footnote{The angular offsets of the overlapping H\ii\ regions disagreed with those reported by \citet{berg2013}. After confirming our H\ii\ region locations with recent study of NGC\,2403 by \citet{mao2018}, we update the positions of the \citet{berg2013} H\ii\ regions to be concurrent with the radial distances of our regions.}. The shapes of the points represent the number of direct temperatures used in the weighted average ionization zone temperatures. Since \citet{berg2013} obtain no direct \siii\ temperatures, the maximum number of temperatures used in the weighted average is two. The oxygen abundance gradient is calculated for these seven regions:
\begin{equation}
\mbox{12+log(O/H)}_{B+13}=8.50(\pm0.09) - 0.07(\pm0.04)R_{g}/R_{e},
\label{eq:bohabun}
\end{equation}
or
\begin{equation}
\mbox{12+log(O/H)}_{B+13}=8.50(\pm0.09) - 0.026(\pm0.016)R_{g}/kpc,
\end{equation}
with an intrinsic dispersion of $\sigma_{int}=0.07\pm0.08$ dex. The redetermined gradient is in agreement with the previously reported gradient for the seven H\ii\ regions, although the magnitude of the intrinsic scatter is larger than previously reported. This scatter is consistent with $\sigma_{int}=0$ dex within uncertainty.

The CHAOS-measured abundance gradient in NGC\,2403 agrees with the \citet{berg2013} redetermined gradient within uncertainty, and the temperatures and oxygen abundances in three of the four overlapping regions agree within uncertainty. We can add the three outer H\ii\ regions of \citet{berg2013}, or those that we have not observed, to our dataset to increase the radial sampling of the H\ii\ regions in NGC\,2403. We measure an abundance gradient for this combined dataset, totaling 30 H\ii\ regions from R$_g$/R$_e$ of 0.39 to 3.59, of
\begin{equation}
\mbox{12+log(O/H)}_{All}=8.56(\pm0.03) - 0.093(\pm0.017)R_{g}/R_{e},
\end{equation}
with an intrinsic dispersion about the gradient of $\sigma_{int}=0.034\pm0.017$ dex. This gradient is consistent with Equations \ref{eq:ohabun} and \ref{eq:bohabun} within statistical uncertainty.

The oxygen abundance determined for the innermost H\ii\ region with direct abundances, NGC\,2403$-$14+42, does not agree with the redetermined abundance. For this region, the low-ionization zone temperature measured, 7800$\pm$300 K, is significantly lower than the redetermined temperature, 8600$\pm$300 K. The difference in the O/H abundances is entirely consistent with the difference in the low-ionization zone temperature: recalculating the abundance within this region using the redetermined low-ionization zone temperature for the O$^+$ abundance (keeping all else constant) yields 12+log(O/H) = 8.46$\pm$0.07 dex, in agreement with the redetermined \citet{berg2013} abundance in this region.

We measure an intrinsic dispersion about the O/H gradient of $\sigma_{int}=0.037\pm0.017$ dex; this is smaller than the redetermined $\sigma_{int}=0.07\pm0.08$ dex from the \citet{berg2013} data, and is not consistent with 0 dex within uncertainty. The same result is found when using the ionization-basedd temperature prioritization method, see Table \ref{t:ipFits} in Appendix \ref{sec:A2}. The sample of H\ii\ regions selected might affect the dispersion in the oxygen abundances: \citet{berg2013} target four bright H\ii\ regions within the first $\sim$3 kpc and three extended H\ii\ regions, while we target many H\ii\ regions within $\sim$4.5 kpc and very few outer H\ii\ regions. The regions we select range from the same bright H\ii\ regions of \citet{berg2013} to a few relatively dim regions that are on the outskirts of the diffuse spiral arms.  As mentioned in \S1, IFU studies have detected abundance enhancement in arm H\ii\ regions relative to interarm regions \citep{ho2019,krec2019,same2020}. If physical processes such as radial mixing are more efficient along the spiral arms of a galaxy, then these processes might be a source of the non-zero scatter in abundances we observe. However, the spiral structure of NGC\,2403 is difficult to trace, and a conservative estimate of the number of interarm H\ii\ regions results in too few regions to make a statistical comparison between the two populations of H\ii\ regions.

\begin{figure}
   \epsscale{1.17}
   \centering
   \plotone{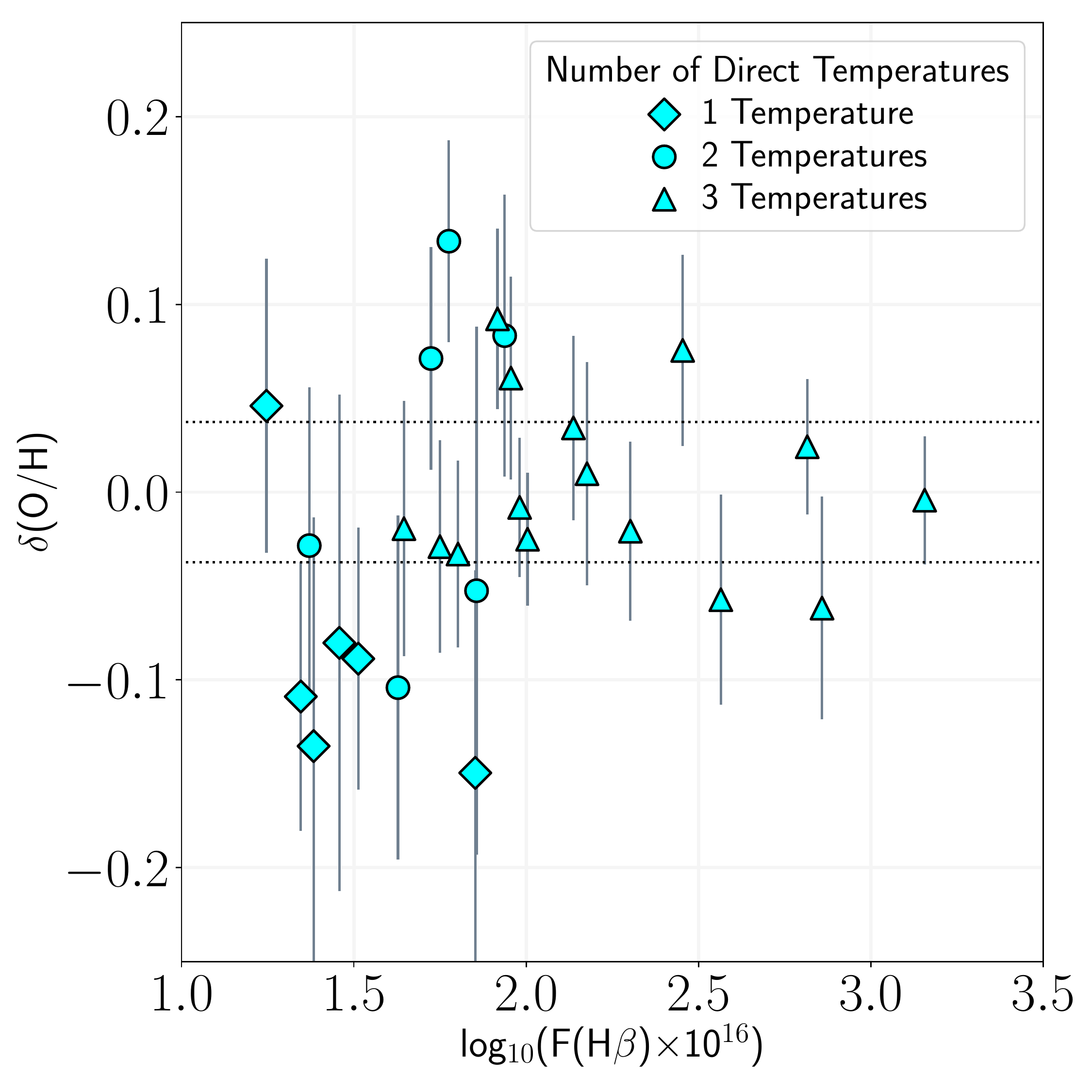}
   \caption{The offset from the O/H gradient in NGC\,2403 vs.\ the observed flux of H$\beta$. The intrinsic dispersion about the gradient is represented by the dotted horizontal lines. The number of direct temperatures from the commonly-used auroral lines (from \nii, \siii, and \oiii) measured in each region is represented by the different shapes. The regions with the most auroral line detections typically fall within the intrinsic scatter about the abundance gradient of Equation \ref{eq:ohabun}, while regions with fewer direct temperatures are more scattered about the gradient.}
   \label{fig:n2403_dOH_FHB}
\end{figure}

To investigate the impact of having more than one temperature measurement, we plot in Figure \ref{fig:n2403_dOH_FHB} the offsets from the best-fit abundance gradient, $\delta$(O/H), vs.\ the flux of H$\beta$ before accounting for stellar absorption. The dotted lines represent the intrinsic dispersion about the abundance gradient in NGC\,2403. The different shapes designate the number of electron temperatures used in the weighted average ionization zone temperature (same convention as the top panel in Figure \ref{fig:ohAbun}). There is no clear trend between offset from the abundance gradient and the flux of H$\beta$. Fitting the average and standard deviation for each population of H\ii\ regions, we find that the H\ii\ regions with the most direct temperatures are most consistent with the gradient ($<$\,$\delta$(O/H)$>$ = 0.00 dex with standard deviation $\sigma$ = 0.04 dex). The regions with two direct temperatures ($<$\,$\delta$(O/H)$>$ = 0.02 dex, $\sigma$ = 0.08 dex) or a single direct temperature ($<$\,$\delta$(O/H)$>$ = $-$0.08 dex, $\sigma$ = 0.06 dex) contain more scatter about the gradient. The regions with the most auroral line detections have the potential to have the lowest oxygen abundance uncertainty due to the weighted average approach to the ionization zone temperatures, so it is likely that the best-fit gradient is weighted to these regions.

Fitting the oxygen abundances in the 15 regions with the \nii, \siii, and \oiii\ auroral line detections, the abundance gradient is measured to be $-0.08\pm0.03$ dex/R$_e$ with an intrinsic scatter of $\sigma_{int} =$ 0.031$\pm$0.018 dex. The intrinsic dispersion is within uncertainty of the dispersion about the gradient in Equation \ref{eq:ohabun} and the dispersion about the gradient obtained when using the ionization-based temperature prioritizations (see Table \ref{t:ipFits}). It should be mentioned that this subsample of H\ii\ regions may have its own biases (for instance, towards bright H\ii\ regions, see Figure \ref{fig:n2403_dOH_FHB}) and that we are not advocating for the rejection of the regions lacking in all available auroral lines. Instead, this reveals that the regions in which the weighted average approach should be performing optimally also have a small, but non-zero intrinsic dispersion about their best-fit gradient. Given these findings, we conclude that the dispersion about the abundance gradient is not a product of the weighted average temperature prioritization method. Perhaps some component is dependent on the use of single temperature H\ii\ regions, but we do not have the large number of H\ii\ regions to examine the true source of the abundance variations in NGC\,2403.

\subsection{CHAOS O/H Gradients}

The bottom panel of Figure \ref{fig:ohAbun} plots the oxygen abundances of the four previously reported CHAOS galaxies in addition to the oxygen abundances of NGC\,2403 against R$_g$/R$_e$. The abundances in the previous galaxies are rederived using the line intensities from \citet{berg2019}, the weighted-average temperature prioritizations, and the ICFs discussed in \S4. Plotting these abundances vs.\ R$_g$/R$_{e}$ allows for a better comparison to IFU abundance studies. While an IFU survey obtains a much larger number of H\ii\ regions and differences between direct and strong-line abundances are to be expected, it is worthwhile to determine if the five CHAOS galaxies have similar oxygen abundance gradients when plotted against R$_g$/R$_{e}$. For clarity, all H\ii\ regions (except for NGC2403+88-18) are represented by the same shapes.

Table \ref{t:abunFits} contains the best-fit gradients and dispersions for the CHAOS galaxies plotted in the bottom panel of Figure \ref{fig:ohAbun}. NGC\,628, NGC\,3184, and NGC\,2403 have abundance gradients within uncertainty of the $-$0.1$\pm$0.03 dex/R$_e$ universal oxygen abundance gradient reported by \citet{sanc2014} and \citet{same2018}. The oxygen abundance gradients are slightly shallower than those reported in \citet{berg2019}, which is the result of the increased uncertainty obtained when applying a T$_e$-T$_e$ relation. The largest change is observed in NGC\,5194's oxygen abundance gradient, which is now slightly positive as opposed to negative. 

The more significant change in the abundance gradient of NGC\,5194 is due to the increased uncertainties on the temperatures and abundances. \citet{croxall2015} detected no \oiii\ auroral lines in the H\ii\ regions of this galaxy, resulting in an inferred temperature for the high-ionization zone in each region. The increased uncertainty on the inferred temperatures propagates into the uncertainty on the abundances, and large uncertainties on all abundance measurements flatten the best-fit gradient. The new gradient reported here for NGC\,5194 is consistent with zero, which is not unexpected for an interacting galaxy \citep[see discussion in][]{croxall2015}. A flatter gradient consistent with zero is observed even when using the ionization-based temperature prioritizations, as seen in Table \ref{t:ipFits} in Appendix \ref{sec:A2}, although the best-fit gradient is negative.

The reported oxygen abundance gradient of NGC\,5457, which is consistent with other studies \citep{kenn2003b,este2020}, is steeper than the universal oxygen abundance gradient observed by the aforementioned IFU studies, and, therefore, is an outlier in that regard. It is also notable that, in the five galaxies, the O/H abundance observed at one half-light radius is fairly constant. The O/H abundance at R$_g$ = R$_e$ ranges from 12+log(O/H) $=$ 8.46 dex in NGC\,2403 to 12+log(O/H) $=$ 8.56 dex in NGC\,628.

Another difference between the findings in this study and \citet{berg2019} is the scatter about the best-fit gradients. Now that the uncertainties on inferred temperatures are larger in magnitude, the uncertainties on the oxygen abundances have increased in the H\ii\ regions that are most reliant on single auroral line detections. Additionally, poor detections are weighted less in the calculation of the ionization zone temperatures, which should result in temperatures/ abundances that are close to the true values within a region. Underestimating the uncertainty on oxygen abundances or using imprecise measurements can result in an overestimation of the intrinsic dispersion about the abundance gradient; this artificially larger scatter could be falsely interpreted as real chemical inhomogeneities within the system \citep[see discussion in][]{este2020}. With the appropriately estimated temperatures and uncertainties, it is expected that the fitted dispersion about the best-fit gradient will decrease in the four previously reported CHAOS galaxies. This is the case for NGC\,628, NGC\,5194, and NGC\,3184, but the intrinsic dispersion of NGC\,5457 is still within the uncertainty of the previously reported value. The intrinsic dispersion values obtained using the ionization-prioritization temperatures with the updated T$_e$-T$_e$ relations are consistent with those reported in Table \ref{t:abunFits} within uncertainty (see Table \ref{t:ipCompare}).

All galaxies have $\sigma_{int}$ significantly above 0 dex. NGC\,2403 has the smallest intrinsic dispersion observed in the non-interacting galaxies, and this dispersion is reproduced when examining the subsample of H\ii\ regions with \oiii, \siii, and \nii\ auroral line detections. This analysis is repeated for NGC\,5457, which is the only other galaxy with a statistically-significant population of H\ii\ regions with all three of these auroral lines detected (26 regions). We find that the intrinsic dispersion about the gradient for these regions drops to $\sigma_{int}$ = 0.052$\pm$0.018 dex, still significantly larger than 0 dex. However, the new gradient is shallower ($-0.15 \pm 0.02$ dex/R$_e$) on account of the shorter radial coverage of these H\ii\ regions: the regions in NGC\,5457 with all three auroral line detections span R$_g$/R$_e$ = 0.86  to 3.55 as opposed to the full sample which covers R$_g$/R$_e$ = 0.43  to 4.58. We do not expect all H\ii\ regions to contain each auroral line, especially the high-metallicity central H\ii\ regions wihere \oiii$\lambda$4363 is difficult to detect. However, this subsample of H\ii\ regions in NGC\,5457 reveals a scatter that is still significantly greater than zero, and so we can conclude that the non-zero intrinsic dispersions in Table \ref{t:abunFits} indicate the presence of non-zero abundance variations in these galaxies.

\subsection{N/O Relative Abundance Gradient}

\begin{figure}
   \epsscale{1.17}
   \centering
   \plotone{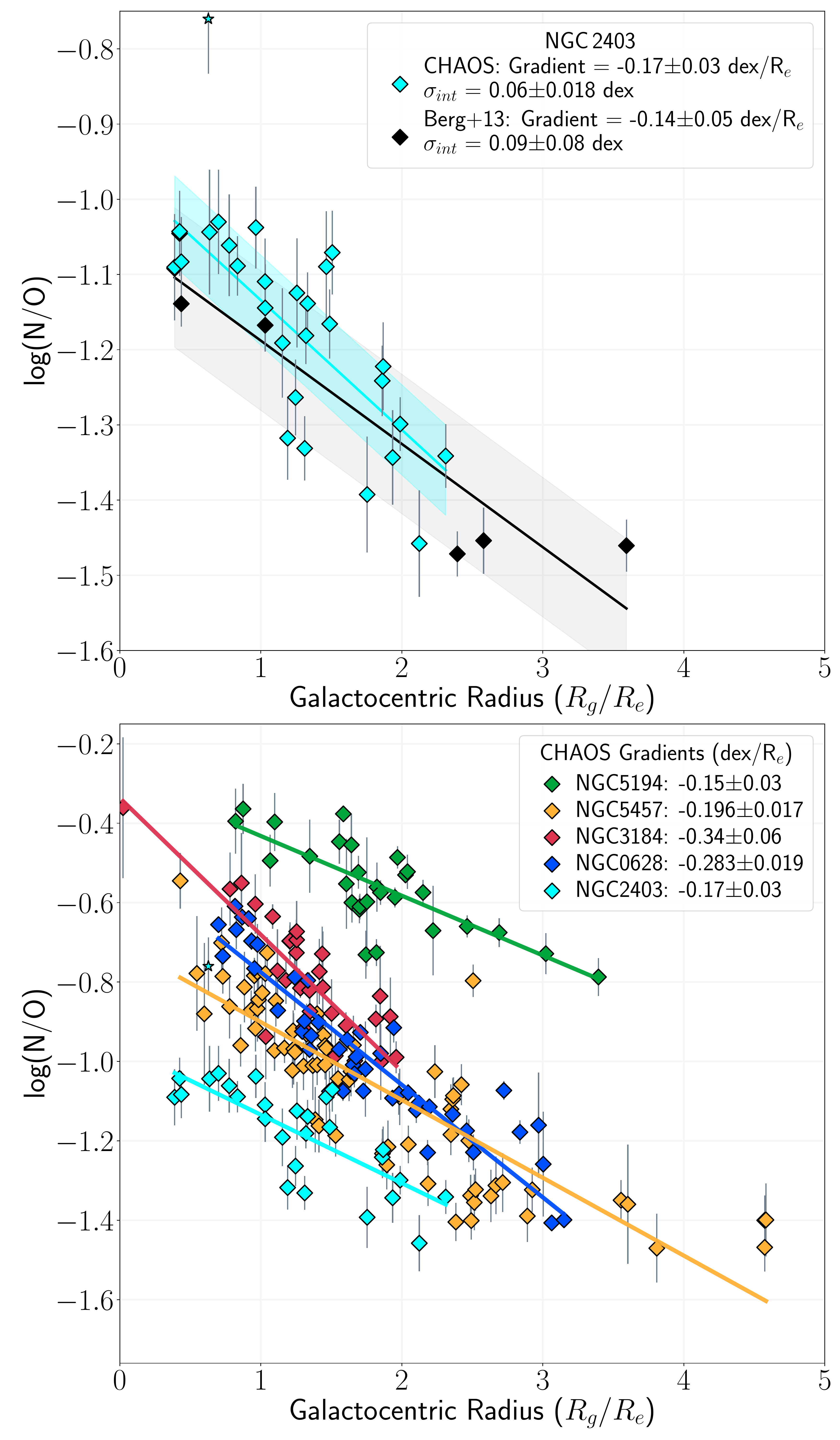}
   \caption{\textit{Top Panel:} CHAOS (cyan) and the recalculated \citet{berg2013}  (black) N/O relative abundances in NGC\,2403 plotted vs.\ R$_g$/R$_e$. The gradients in each galaxy are plotted as solid lines and are provided in the legend. The intrinsic dispersion about each gradient, represented as the shaded region around each gradient, are also reported in the legend. \textit{Bottom Panel:} Same as top panel but for the N/O relative abundances of all CHAOS galaxies. The shaded portions around each gradient are removed for clarity, and the galaxies listed in the legend are ordered by decreasing stellar mass.}
   \label{fig:n2403_noAbun}
\end{figure}

Nitrogen has both primary and secondary origins \citep{henr2000}. N/O gradients trace intermediate-mass stars' contribution to secondary N relative to the primary N and O production in high-mass stars. The trend observed in previous CHAOS galaxies is a negative N/O radial gradient with small scatter relative to the scatter observed in O/H. The N/O relative abundances of NGC\,2403 from this study and those recalculated from the line intensities of \citet{berg2013} are plotted against R$_g$/R$_{e}$ in the top panel of Figure \ref{fig:n2403_noAbun}; the N/O abundances of all CHAOS galaxies are plotted in the bottom panel of this figure. The best-fit N/O gradients in the top panel are consistent within uncertainty:
\begin{align}
\mbox{log(N/O)}= -0.96(\pm0.05) - 0.17(\pm0.03)R_{g}/R_{e} \\
\mbox{log(N/O)}_{B+13}=-1.05(\pm0.09) - 0.14(\pm0.05)R_{g}/R_{e}.
\end{align}
As described in \S5.1, NGC2403+88$-$18, the region represented by a cyan star, is not included in the fit. The intrinsic dispersion about the fits are 0.060$\pm$0.018 dex and 0.09$\pm$0.08 dex, respectively. Including the outer 3 H\ii\ regions of \citet{berg2013} in our dataset results in an N/O gradient of $-0.16\pm0.02$ dex/R$_e$, consistent with the above results. Additionally, the N/O abundances of all the overlapping H\ii\ regions agree within uncertainty.

%\begin{figure*}[!t] 
%\epsscale{0.7}
\begin{figure} 
\epsscale{1.17}
   \centering
   \plotone{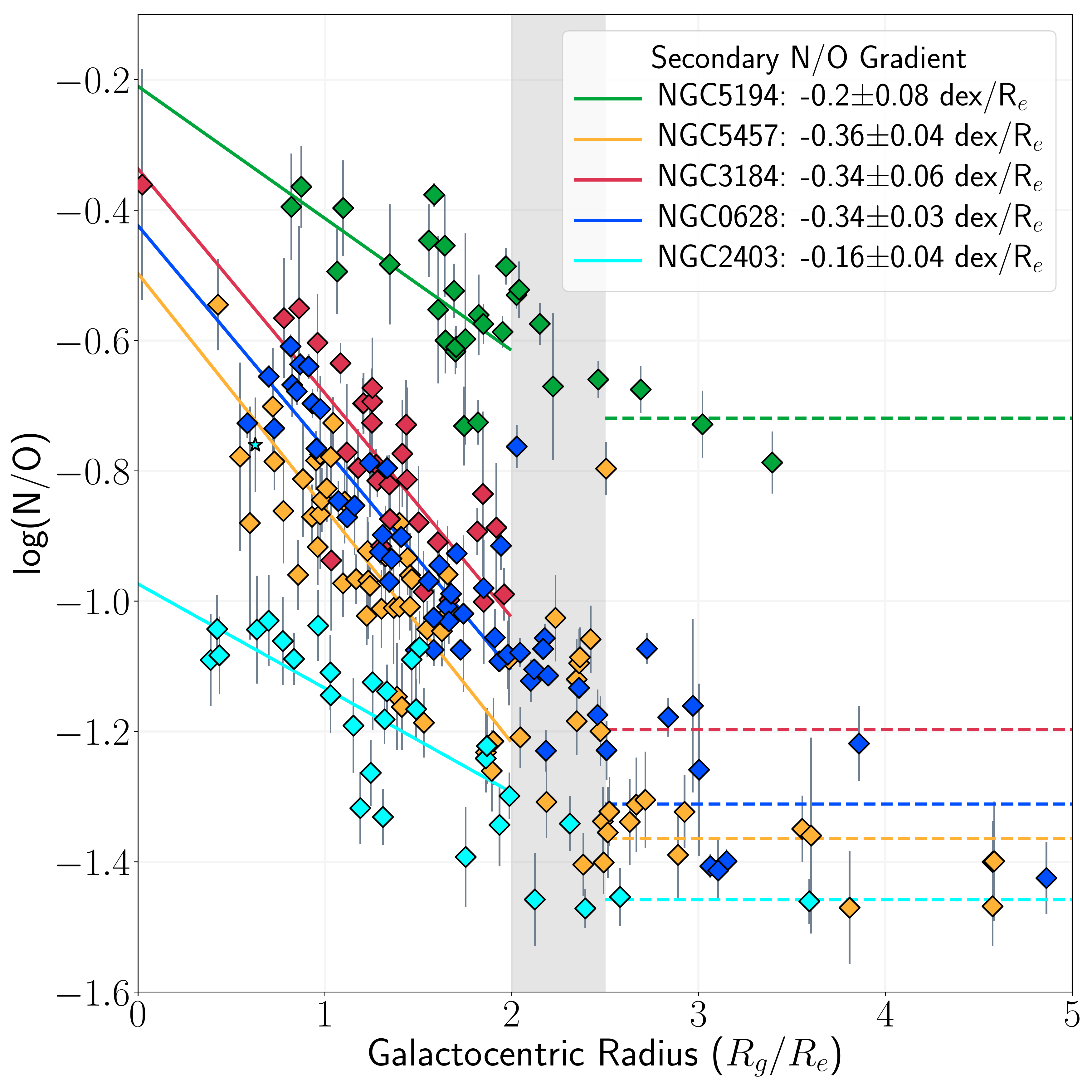}
   \caption{N/O relative abundances in the five CHAOS galaxies plotted vs.\ R$_g$/R$_e$. The secondary N/O gradients in each galaxy are plotted as solid lines from R$_g$/R$_e$ $=$ 0 to 2.0, and the primary N/O plateaus are plotted as dashed lines from R$_g$/R$_e$ $=$ 2.5 and beyond. The secondary N/O gradients are provided in the legend. The grey shaded box indicates the region where primary N/O transitions to secondary, and the regions within this area are not used in the fits. The secondary N/O gradients of NGC\,5457, NGC\,3184, and NGC\,628 are all consistent, while NGC\,2403's secondary N/O gradient is significantly shallower.}
   \label{fig:chaos_noPrim}
%\end{figure*} 
\end{figure} 

The central H\ii\ region contains a relative N/O that agrees with the redetermined value, despite having a significantly lower low-ionization zone temperature. The emissivity of \nii$\lambda$6584 relative to that of \oii$\lambda$3727 is dependent on T$_e$ such that j$_{\textrm{[N II]}6584}$/j$_{\textrm{[O II]}3727}$ decreases with increasing T$_e$. As T$_e$ increases to higher temperatures, the proportional change in the relative emissivities becomes smaller. The difference in the emissivity using the temperature reported here and the temperature redetermined from the previous line intensities is $\sim$20\%. However, a similar difference in the relative intensities of \nii$\lambda$6584 and \oii$\lambda$3727 is found when comparing the line intensities measured here (see Table \ref{t:intensities}) to those in Table 7 of \citet{berg2013}. The net result is an agreement in the relative N/O abundances, while the O/H abundances are discrepant due to the difference in low-ionization zone temperatures.

\citet{berg2019} found evidence for a universal secondary N/O gradient of $-$0.33 dex/R$_e$ at R$_{g}$/R${_e}$ $<$ 2.0 in non-interacting galaxies, implying that secondary N production dominates at these radii. This result is obtained by fitting the combined N/O abundances at R$_{g}$/R${_e}$ $<$ 2.0 with a single gradient and using the y-intercept or each galaxy's primary N/O plateau at R$_{g}$/R${_e}$ $>$ 2.5 to scale each of the non-interacting galaxies. The most distant H\ii\ region that we observe in NGC\,2403 is located at R$_{g}$/R${_e}$ = 2.31, but we use the combined dataset with the recalculated \citet{berg2013} N/O abundances in the outer 3 H\ii\ regions to obtain an estimate on the primary N/O plateau in NGC\,2403. Figure \ref{fig:chaos_noPrim} plots the N/O relative abundances, secondary N/O gradient, and primary N/O plateau in each galaxy. Secondary N/O gradients are obtained using LINMIX. The primary N/O plateaus are calculated using a weighted average of the N/O abundances in the regions at R$_g$/R$_e$ $>$ 2.5 in each galaxy. The regions found at 2.0 $<$ R$_g$/R$_e$ $<$ 2.5 are considered to be in transition from primary to secondary N production and are excluded from the fits. We note that NGC\,3184 contains no regions at R$_g$/R$_e$ $>$ 2.5, requiring an extrapolation of this galaxy's secondary N/O gradient to estimate the primary N/O plateau. We also note that \citet{berg2019} also use the H\ii\ regions observed in NGC\,628 by \citet{berg2013} to measure this galaxy's primary N/O plateau. Consistent with our approach to the regions of NGC\,2403, we recalculate the N/O abundances in the \citet{berg2013} regions of NGC\,628 in a method consistent with those described in \S3 and \S4 before combining them with the CHAOS NGC\,628 data.

The secondary N/O gradient at R$_{g}$/R${_e}$ $<$ 2.0 in NGC\,2403 is
\begin{equation}
\mbox{log(N/O)}_{Sec}=-0.97(\pm0.05) - 0.16(\pm0.04)R_{g}/R_{e},
\end{equation}
and the primary N/O is $-$1.46 dex. If we determine the N/O abundance at R$_{g}$/R${_e}$ $=$ 2.5 via the above equation (similar to the approach taken for NGC\,3184), we measure a primary N/O plateau of $-$1.37 dex, similar to the primary N/O plateau measured in NGC\,5457. We do not expect exact agreement amongst the primary N/O plateaus of the CHAOS galaxies, as primary N and O production is dependent on the star formation history in the galaxy \citep[see discussion in][]{berg2019}. However, the secondary N/O gradient in NGC\,2403 is significantly shallower than the other non-interacting CHAOS galaxies, which indicates that the secondary N production in NGC\,2403 is unique in this sample of galaxies. The stellar mass of NGC\,2403 is also smaller than the other CHAOS galaxies. The stellar masses for the CHAOS galaxies are log(M$_{\star}$/M$_\odot$) = 9.6, 10.0, 10.3, 10.4, and 10.7 for NGC\,2403, NGC\,628, NGC\,3184, NGC\,5457, and NGC\,5194, respectively.\footnote{Z0MGS WISE observations presented in \citet{lero2019} are used to obtain the stellar masses. These masses are dependent on the luminosity and, therefore, the adopted distance to each galaxy. We have scaled the stellar masses of each CHAOS galaxy to the distances reported in Table 1 of \citet{berg2019} and in Table \ref{t:n2403global}.} NGC\,2403, with its smaller stellar mass, is less chemically evolved, and so secondary nitrogen production has not had time to establish a strong N/O gradient in the galaxy.

The intrinsic dispersion about the observed N/O gradient in NGC\,2403 is consistent with the dispersion observed within the other galaxies. The intrinsic dispersions about the best-fit N/O gradients in NGC\,5194, NGC\,5457, and NGC\,3184 have increased from the previous results in \citet{berg2019}. This is because the errors on N/O have decreased for all regions due to the weak dependence of j$_{\textrm{[N II]}6584}$/j$_{\textrm{[O II]}3727}$ on the low-ionization zone electron temperature at T$_e$ $>$ 8000 K. As such, the dominant source of uncertainty is in the measured fluxes of the strong nebular lines, which is small for the average CHAOS spectrum. Additionally, the single N/O gradient is not an appropriate fit for some of the galaxies; the secondary N/O gradient with primary N/O plateau provides a better fit to the relative abundances in galaxies like NGC\,5457 and NGC\,628 (see Figure \ref{fig:chaos_noPrim}). The dispersion about the N/O gradient in NGC\,2403 is dominated by five regions at low N/O relative to the gradient between R$_g$/R$_e =$ 1 and $\sim$2. Interestingly, these regions are found within or nearby the same diffuse spiral arm.

\section{\texorpdfstring{$\alpha$}{a} Element Abundance Trends}

\begin{figure*}
   \epsscale{1.2}
   \centering
   \plotone{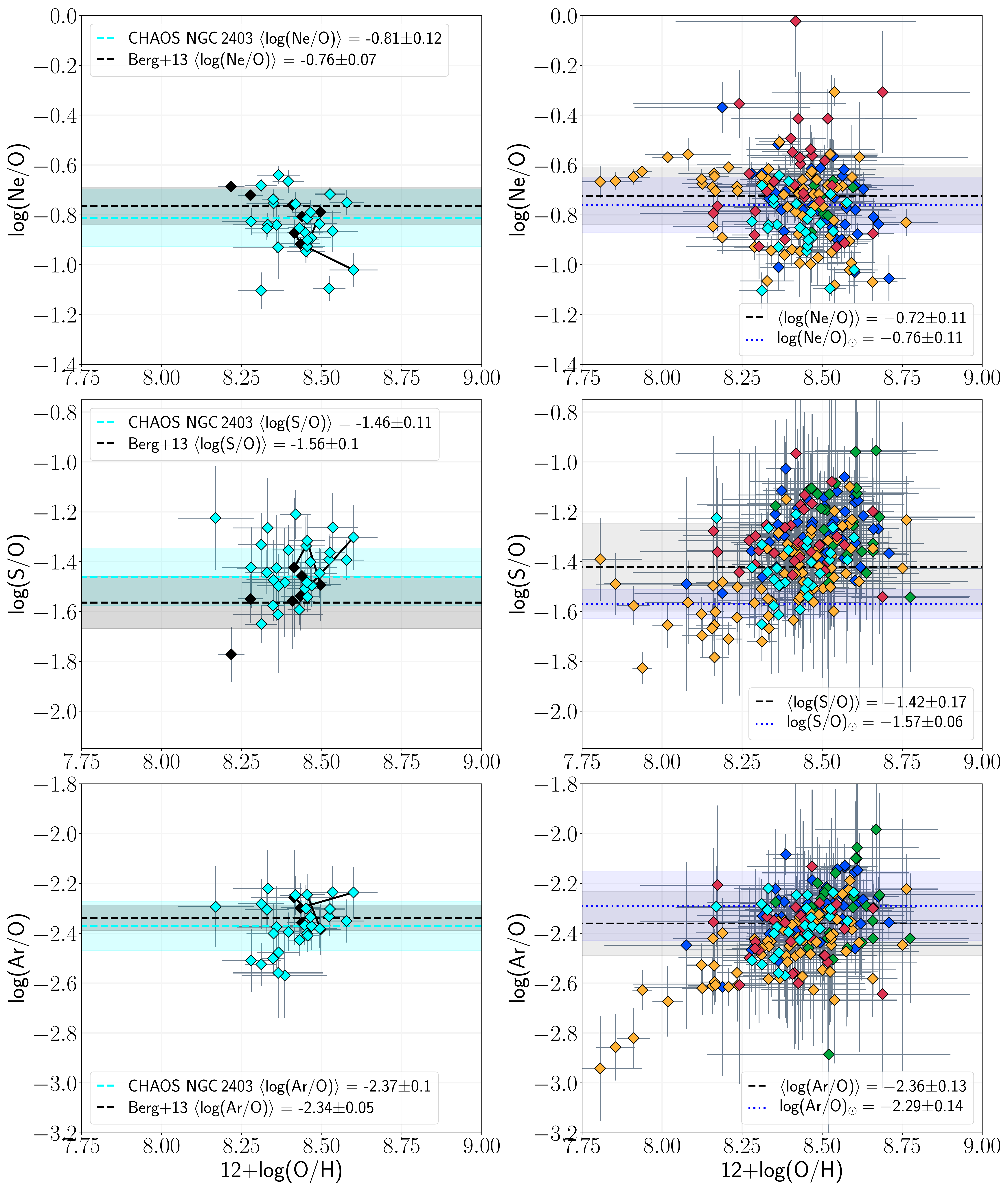}
   \caption{$\alpha$/O relative abundances of the CHAOS galaxies plotted vs.\ 12+log(O/H). \textit{Left Column}: Cyan points are $\alpha$/O abundances in NGC\,2403 as observed by CHAOS, while black points are the redetermined $\alpha$/O relative abundances from \citet{berg2013}. The weighted averages and 1$\sigma$ uncertainty of each dataset are plotted as dashed lines and shaded regions, respectively. The solid black lines connecting some of the CHAOS and \citet{berg2013} data represent overlapping H\ii\ regions. \textit{Right Column}: All CHAOS $\alpha$/O data for the five galaxies (color-coding same as Figure \ref{fig:chaosTTFig}). The black dashed line and the grey shaded portion is the best-fit weighted average $\alpha$/O and its 1$\sigma$ uncertainty, respectively. The dotted blue line and shaded blue area is the solar $\alpha$/O value and its uncertainty, respectively, from \citet{aspl2009}.}
   \label{fig:chaos_ao}
\end{figure*}

Production of different $\alpha$ elements takes place in the same progenitors (high-mass stars), and so it is expected that the relative abundance of these elements is fairly constant at a given metallicity. It is common to compare the relative abundances of $\alpha$/O to O/H to see if the relative production of an $\alpha$ element varies as a function of metallicity. The Ne/O, S/O, and Ar/O relative abundances of NGC\,2403 and the full CHAOS sample vs.\ O/H are plotted in Figure \ref{fig:chaos_ao}. The left column examines the $\alpha$/O trends observed in NGC\,2403 and compares these abundances to the redetermined \citet{berg2013} $\alpha$ element abundances. The dashed lines and shaded regions correspond to the weighted average and 1$\sigma$ uncertainty of each dataset, respectively. Additionally, the overlapping regions are connected by solid black lines for easier comparisons between the common H\ii\ regions. The right column plots all CHAOS $\alpha$/O data. In these panels, the blue dotted line is the solar value of the plotted $\alpha$/O ratio \citep[from][]{aspl2009}, the light blue shaded box is the uncertainty on the solar average, the black dashed line is the weighted average of the CHAOS data, and the grey box about the CHAOS average is the 1$\sigma$ error on the average.

We first address the NGC\,2403 $\alpha$ element abundances in the left column. There is generally good agreement between the redetermined $\alpha$ element abundances from the intensities of \citet{berg2013} and those determined in this study. The average values determined in NGC\,2403 are log(Ne/O)$_{avg}$ = $-$0.81$\pm$0.12 dex, log(S/O)$_{avg}$ = $-$1.46$\pm$0.11 dex, and log(Ar/O)$_{avg}$ = $-$2.37$\pm$0.10 dex. The averages determined for the updated \citet{berg2013} data are log(Ne/O)$_{B+13}$ = $-$0.76$\pm$0.07 dex, log(S/O)$_{B+13}$ = $-$1.56$\pm$0.10 dex, and log(Ar/O)$_{B+13}$  = $-$2.34$\pm$0.05 dex. Both the average Ne/O and Ar/O values of each sample agree within 0.04 dex, and both samples exhibit no Ne/O trend as a function of metallicity. The Pearson's correlation coefficient for the CHAOS NGC\,2403 Ar/O vs.\ O/H data is 0.31 with a p value of 0.10, suggesting that there might be a weak, positive correlation of Ar/O as a function of O/H in these regions. No such trend is observed in the S/O data (Pearson correlation coefficient of 0.09 with a p value of 0.64 for the CHAOS NGC\,2403 H\ii\ regions). The CHAOS NGC\,2403 average S/O is larger than the updated \citet{berg2013} average S/O by 0.11 dex, almost outside the uncertainties of either average.

When considering the differences in how sulfur abundances are calculated, the discrepancy between log(S/O)$_{avg}$ and log(S/O)$_{B+13}$ is not altogether surprising. The source of the discrepancy is the use of the \siii$\lambda$6312 emissivity in the determination of the \citet{berg2013} S$^{2+}$ abundances. \citet{berg2013} used the Blue Channel Spectrograph on the MMT to obtain their H\ii\ region spectra; this spectrograph has a wavelength cutoff at 6790 \AA, which means the strong nebular \siii\ lines are not used for abundance determination. The emissivity of \siii$\lambda$6312 is a strong function of the electron temperature, hence it is useful as a temperature diagnostic. However, this sensitivity, coupled with the use of inferred temperatures from T$_e$-T$_e$ relations, makes abundances using the emissivity of \siii$\lambda$6312 strongly dependent on the physical conditions within a different ionization zone. The weighted average of the inferred temperatures from the recalculated T$_e$\nii\ and T$_e$\oiii\ is not enough to resolve the issue of using \siii$\lambda$6312 for S$^{+2}$ abundance determination. Given the wavelength coverage of MODS, we use the strong \siii$\lambda\lambda$9069,9532 nebular lines, corrected to the theoretical ratio if there is evidence of contamination (as described at the beginning of \S3), for abundance determination. These lines have a weaker dependence on the electron temperature and are at significantly higher S/N than \siii$\lambda$6312.

We now turn our attention to the net $\alpha$ element abundances in the right column of Figure \ref{fig:chaos_ao}. The solar values are in agreement with the CHAOS weighted averages for all three $\alpha$ elements, within statistical uncertainties. The weak temperature dependence of the relative [\ion{Ne}{3}]$\lambda$3686 and \oiii$\lambda$5007 emissivities result in small uncertainties on the Ne/O relative abundances. This is similar to the small N/O uncertainties discussed in \S5.3, although the ratio of the \nii$\lambda$6584 and \oii$\lambda$3727 emissivities is a stronger function of T$_e$ at low electron temperatures. The dispersion in Ne$^{2+}$/O$^{2+}$ at low ionization manifests as increased scatter below the best-fit average at intermediate and high metallicity. Despite this dispersion, the average of the CHAOS data agree with the solar Ne/O. While there is no S/O trend as a function of O/H in the NGC\,2403 data, the Pearson correlation coefficient for the net CHAOS dataset is 0.46. The evidence for increasing S/O with O/H is primarily found in NGC\,5457, although regions with the lowest metallicity do not follow the trend. Additionally, the Pearson correlation coefficient for Ar/O vs.\ O/H is 0.54, indicating that the trend is more evident for all the regions than just those of NGC\,2403. \citet{este2020} find a trend of increasing Ar/O vs.\ O/H in NGC\,5457. With the updated temperatures and ICF \citep[which now matches the ICF used by][]{este2020}, the CHAOS data do appear to increase in Ar/O at increasing O/H. The lack of H\ii\ regions with 12+log(O/H) $<$ 8.2 and the spread in Ar/O in the majority of the H\ii\ regions limit our ability to further explore this trend.

%%%%%%%%%%%%%%%%%%%%%%%%%%%%%

\section{Conclusions}

% Abundance Gradient Table
\begin{deluxetable*}{llccccc}[H]
\tablecaption{CHAOS Abundance Fits}
\tablehead{
\CH{$y$}  & \CH{$x$}  & \CH{Galaxy} & \CH{$\#$ Reg.} & \CH{Equation} & \CH{$\sigma_{\rm{int.}}$} & \CH{$\sigma_{\rm{tot.}}$}}
\startdata
12+log(O/H) (dex)& $R_g$ ($R_e^{-1}$)  &{NGC~
0628} & 45 & $y=(8.65\pm0.05) - (0.09\pm0.03)\times x$ & $0.082\pm0.015$ & 0.099 \\
                &                     & {NGC~5194} & 28 & $y=(8.48\pm0.10) + (0.03\pm0.05)\times x$ & $0.04\pm0.02$ & 0.07 \\
                &                     & {NGC~5457} & 71 & $y=(8.70\pm0.04) - (0.172\pm0.016)\times x$ & $0.097\pm0.013$ & 0.113 \\
                &                     & {NGC~3184} & 30 & $y=(8.63\pm0.15) - (0.14\pm0.10)\times x$ & $0.08\pm0.03$ & 0.11 \\
                &                     & {NGC~2403} & 27 & $y=(8.55\pm0.04) - (0.09\pm0.03)\times x$ & $0.037\pm0.017$ & 0.063 \\
\vspace{-1ex} \\             
log(N/O) (dex)  & $R_g$ ($R_e^{-1}$)  & {NGC~0628} & 45 & $y=(-0.49\pm0.03) - (0.283\pm0.019\times x$ & $0.072\pm0.010$	& 0.075	\\
                &                     & {NGC~5194} & 28 & $y=(-0.28\pm0.06) - (0.15\pm0.03)\times x$ & $0.072\pm0.015$	& 0.075	\\
                &                     & {NGC~5457} & 71 & $y=(-0.70\pm0.03) - (0.196\pm0.017)\times x$ & $0.110\pm0.012$	& 0.120	\\
                &                     & {NGC~3184} & 30 & $y=(-0.34\pm0.08) - (0.34\pm0.06)\times x$ & $0.059\pm0.017$	& 0.076  \\
                &                     & {NGC~2403} & 27 & $y=(-0.96\pm0.05) - (0.17\pm0.03)\times x$ & $0.060\pm0.018$	& 0.075  \\
\vspace{-1ex} \\
log(Ne/O) (dex) &           & {ALL} & 166  & $y=-0.72\pm0.11$                             &      &  \\
\vspace{-1ex} \\
log(S/O) (dex)&               & {ALL} &  202 & $y=-1.42\pm0.17$                             &      &   \\
\vspace{-1ex} \\
log(Ar/O) (dex) &            & {ALL} &  201  & $y=-2.36\pm0.13$                            &      &
\enddata
\tablecomments{
Best-fit equations for the oxygen, nitrogen, and $\alpha$/O abundances observed in the CHAOS galaxies. The first and second columns are the dependent and independent variables, respectively, used for the Equation listed in the fifth column. The galaxy and the number of regions used for the fit are given in the third and fourth column. The intrinsic and total scatter, both in dex, are listed in the sixth and seventh columns, respectively.}
\label{t:abunFits}
\end{deluxetable*}

The CHemical Abundances of Spirals (CHAOS) project has observed nearby, face-on spiral galaxies to build a catalogue of high-quality H\ii\ region spectra for direct electron temperature measurements. The results of the fifth CHAOS galaxy, NGC\,2403, are presented here. With the addition of NGC\,2403, the CHAOS database has grown to include 213 H\ii\ regions with more than one temperature-sensitive auroral line detection.

With a large database of H\ii\ regions with multiple direct temperatures, we create statistically significant empirical T$_e$-T$_e$ relations to infer electron temperatures in different ionization zones. The \citet{croxall2016} empirical T$_e$-T$_e$ relations, developed using the large, homogeneous dataset of H\ii\ regions in NGC\,5457, provide no prescription for estimating errors on the inferred electron temperature, nor are the intrinsic dispersions about the relations used. We propose that the intrinsic dispersion in T$_{e}$ is due to real differences in the physical properties of an H\ii\ region that lead to variations in electron temperature within an ionization zone.

The newly measured intrinsic dispersions are used to, partially, account for these unknowns when applying a T$_e$-T$_e$ relation via the uncertainty on the inferred temperatures. When using an empirical, linear T$_e$-T$_e$ relation, we propose adding the intrinsic dispersion in quadrature with the uncertainty on the measured temperature to obtain more realistic uncertainties on the inferred electron temperature (see Equation \ref{eq:dT_eq}). This sets a minimum uncertainty on inferred electron temperatures, which is appropriate to account for physical differences between ionization zones within an H\ii\ region.

With updated temperatures and uncertainties, we recompute the abundances of all H\ii\ regions from \citet{berg2019} along with the 28 H\ii\ regions in NGC\,2403 with \nii\/, \siii\/, or \oiii\ auroral line detections. We apply a new method to determine the temperatures in each ionization zone: the temperature used is the weighted average of all temperature data in the region, including the dominant ion temperature and inferred ion temperatures when available. The ionization-based temperature prioritizations described in \citet{berg2019} produce similar results (see Appendix \ref{sec:A2}). The ICFs adopted for abundance determinations are slightly different from those previously applied in other CHAOS studies, as discussed in \S4. The oxygen abundance gradients of NGC\,628, NGC\,5457, and NGC\,3184 are consistent with those previously reported in \citet{berg2019}, within uncertainty. The new gradient for NGC\,5194 is consistent with zero, which is not unexpected for an interacting galaxy.

To compare the results for NGC\,2403 to the literature values, we use the line intensities from \citet{berg2013} and recalculate the temperatures and abundances in a method consistent with the one applied to all CHAOS galaxies. The oxygen abundance gradient in NGC\,2403 agrees with the redetermined \citet{berg2013} gradient. Additionally, the intrinsic dispersion about this gradient is the smallest observed in the sample of five galaxies. The intrinsic dispersion in O/H measured in each CHAOS galaxy is significantly greater than zero, and we associate the dispersion in NGC\,2403 with real abundance variations in the galaxy.

The N/O abundances in NGC\,2403 agree with the \citet{berg2013} abundances: all four H\ii\ regions that were previously observed have similar N/O, and the N/O gradients agree within uncertainty. The secondary N/O gradient in NGC\,2403 is the shallowest of all non-interacting CHAOS galaxies; the low stellar mass of NGC\,2403 might indicate that this galaxy has not produced a sufficient amount of secondary nitrogen to establish a steep N/O gradient.

There is agreement between the NGC\,2403 Ne/O and Ar/O abundances reported here and recalculated from the \citet{berg2013} line intensities. The agreement between the S/O abundances is not as clear, but this is due to the different emission lines used for S$^{2+}$ abundance calculation. The $\alpha$/O best-fit averages for the entire CHAOS dataset agree with the solar values, within statistical uncertainty. We observe a possible trend of increasing S/O and Ar/O with O/H in the CHAOS data, a trend that is also reported by \citet{este2020}. More H\ii\ region abundance data, particularly at low O/H, are needed to verify this trend. With more H\ii\ region data, we can begin to develop robust empirical ICFs that appropriately cover the full range of ionization observed in H\ii\ regions and better fit the empirical sulfur, neon, and argon data.

% Update later
\acknowledgments
%Funding
We are grateful to the referee for their insightful comments and thorough feedback which have substantially improved the clarity and depth of this paper.

This work has been supported by NSF Grants AST-1109066 and AST-1714204. This paper uses data taken with the MODS spectrographs built with funding from 
NSF grant AST-9987045 and the NSF Telescope System Instrumentation Program (TSIP), with additional funds from the Ohio Board of Regents and the Ohio State University Office of Research. 
This paper made use of the modsIDL spectral data reduction pipeline 
developed by KVC in part with funds provided by NSF Grant AST-1108693.
This work was based in part on observations made with the Large Binocular Telescope (LBT). 
The LBT is an international collaboration among institutions in the United States, Italy and Germany. 
The LBT Corporation partners are: the University of Arizona on behalf of the Arizona university system; 
the Istituto Nazionale di Astrofisica, Italy; the LBT Beteiligungsgesellschaft, Germany, representing the 
Max Planck Society, the Astrophysical Institute Potsdam, and Heidelberg University; the Ohio State 
University; and the Research Corporation, on behalf of the University of Notre Dame, the University 
of Minnesota, and the University of Virginia. 

%Jiayi's Funding
The work of J.S. is partially supported by the National Science Foundation (NSF) under Grants No. 1615105, 1615109, and 1653300, as well as by the National Aeronautics and Space Administration (NASA) under ADAP grants NNX16AF48G and NNX17AF39G.

%IDL Builders
We are grateful to D. Fanning, J.\,X. Prochaska, J. Hennawi, C. Markwardt, and M. Williams, and others 
who have developed the IDL libraries of which we have made use: coyote graphics, XIDL, idlutils, MPFIT, MPFITXY, and impro.  

%NED
This research has made use of the NASA/IPAC Extragalactic Database (NED) which is 
operated by the Jet Propulsion Laboratory, California Institute of Technology, 
under contract with the National Aeronautics and Space Administration.
%2MASS Acknowledgements
This publication makes use of data products from the Two Micron All Sky Survey (2MASS), which is a joint project of the University of Massachusetts and the Infrared Processing and Analysis Center/California Institute of Technology, funded by the NASA and the NSF.

\clearpage

%%%%%% APPENDIX %%%%%%
\appendix
%Added to change the names of the Tables and Figures after the Conclusion to A#
%\beginsupplement

\section{NGC 2403 Line Intensities, Temperatures, and Abundances}
\label{sec:A1}

\renewcommand{\thetable}{A.1}

%Line Intensity Tables
\LongTables
\begin{deluxetable*}{lcccccccc}[H]
\tabletypesize{\scriptsize}
\setlength{\tabcolsep}{0.03in} 
\tablecaption{Emission Line Intensities in NGC\,2403}
\tablewidth{0pt}
\tablehead{\CH{} & \multicolumn{7}{c}{$I(\lambda)/I(\mbox{H}\beta)$}}
\startdata
{Ion}              & +19-22                & -23-16                & -14+42                & -38+51                & +7+37                 & -27-28                & +56-59                \\
\hline
{H14~\W3721}       &  0.014$\pm$0.004         &  0.025$\pm$0.003         &  0.023$\pm$0.002         &  0.020$\pm$0.002         &  0.023$\pm$0.002         &  0.023$\pm$0.003         &  0.014$\pm$0.003         \\
{[O~\ii]~\W3727}   &  1.591$\pm$0.105         &  2.509$\pm$0.178         &  2.412$\pm$0.178         &  2.030$\pm$0.135         &  2.125$\pm$0.147         &  2.349$\pm$0.074         &  1.712$\pm$0.105         \\
{H13~\W3734}       &  0.017$\pm$0.005         &  0.031$\pm$0.004         &  0.029$\pm$0.003         &  0.025$\pm$0.002         &  0.029$\pm$0.002         &  0.028$\pm$0.003         &  0.017$\pm$0.004         \\
{H12~\W3750}       &  0.034$\pm$0.004         &  0.033$\pm$0.007         &  0.034$\pm$0.003         &  0.043$\pm$0.008         &  0.041$\pm$0.005         &  0.045$\pm$0.007         &  0.021$\pm$0.007         \\
{H11~\W3770}       &  0.032$\pm$0.010         &  0.046$\pm$0.009         &  0.046$\pm$0.005         &  0.049$\pm$0.008         &  0.049$\pm$0.004         &  0.045$\pm$0.005         &  0.019$\pm$0.008         \\
{H10~\W3797}       &  0.038$\pm$0.011         &  0.066$\pm$0.009         &  0.062$\pm$0.005         &  0.056$\pm$0.005         &  0.064$\pm$0.005         &  0.062$\pm$0.007         &  0.037$\pm$0.008         \\
{He~I~\W3819}      &  0.011$\pm$0.012         &  0.007$\pm$0.010         &  0.013$\pm$0.002         &  \nodata          &  0.005$\pm$0.003         &  0.004$\pm$0.008         &  \nodata          \\
\enddata
\tablecomments{The reddening-corrected line intensities in the H\ii\ regions of NGC\,2403. These intensities are relative to I(H$\beta$). $C(H\beta)$, $a_H$ (in \AA), and the flux of H$\beta$ before correction for stellar absorption (in 10$^{-16}$ergs s$^{-1}$ cm$^{-2}$) are provided in the last three rows.}
\label{t:intensities}
\end{deluxetable*}

%------------------------------------------------------------------
\clearpage

\renewcommand{\thetable}{A.2}

\begin{deluxetable*}{lcccccccc}
\tablecolumns{8}
\tablewidth{0pt}	
\tabletypesize{\scriptsize}
\setlength{\tabcolsep}{0.03in} 
\tablecaption{Temperatures, Ionic and Total Abundances in NGC\,2403}
\startdata
\hline\hline \\
Property                   & -14+42                & -38+51                & +7+37                 & +88-18                & -97+39                & -84-0                 & -3-71                 \\
\hline
$T_e$[S\ii]~(K)             &  10500$\pm$1100       &  8000$\pm$800       &  8500$\pm$500       &  12000$\pm$1200       &  \nodata             &  \nodata             &  \nodata             \\
$T_e$[N\ii]~(K)             &  7700$\pm$700       &  7900$\pm$300       &  8000$\pm$300       &  \nodata             &  \nodata             &  \nodata             &  \nodata             \\
$T_e$[O\ii]~(K)             &  9500$\pm$500       &  9500$\pm$400       &  9300$\pm$400       &  10100$\pm$600       &  9300$\pm$500       &  12500$\pm$700       &  10700$\pm$1100       \\
$T_e$[S\iii]~(K)            &  7300$\pm$300       &  7900$\pm$400       &  7600$\pm$300       &  8200$\pm$900       &  8500$\pm$1000       &  8000$\pm$400       &  8000$\pm$800       \\
T$_e$[O~\iii] (K)          &  \nodata             &  8800$\pm$300       &  9200$\pm$400       &  \nodata             &  \nodata             &  \nodata             &  \nodata             \\
$n_{e}$~(cm$^{-3}$)       &  50$\pm$100       &  120$\pm$120       &  60$\pm$100       &  30$\pm$80       &  40$\pm$110       &  130$\pm$100       &  10$\pm$70       \\
\enddata
\tablecomments{Electron temperatures and ionic abundances in the H\ii\ regions of NGC\,2403. All regions contain at least one detection of the \nii, \siii, or \oiii\ auroral lines. The temperatures in each ionization zone are calculated using the weighted average of the available direct temperatures and the T$_e$-T$_e$ relations of \S3.2.} 
\label{t:abundances}
\end{deluxetable*}

\section{Ionization-Based and Weighted Average Temperature Prioritizations}\label{sec:A2}

\citet{berg2019} use ionization-based temperature prioritizations to determine the electron temperature in an ionization zone. These prioritizations are constructed to make use of electron temperatures from dominant ionization zones within an H\ii\ region. Here, we compare the ionization-based temperature prioritizations to the weighted-average temperatures within the ionization zones described in \S4.

The temperatures prioritized in the ionization-based approach are dependent on the $O_{32} = \frac{I(5007)}{I(3727)}$ parameter within a region. The schematic for the ionization-based prioritizations is provided in Figure \ref{fig:temp_priors}, where the equations are the T$_e$-T$_e$ relations discussed in \S3.2 \citep[this figure has been modified from Figure 5 in][]{berg2019}. The low-ionization prioritizations are the same for a region with high or low average ionization. A direct \siii\ temperature is prioritized in the intermediate-ionization zone when available, independent of the average ionization. If \siii$\lambda$6312 is not detected, then the temperature from either \oiii\ or \nii\ is used, depending on the ionization of the region.

However, there are substantial differences in the prioritization of electron temperatures for the high-ionization zone. In H\ii\ regions characterized by high average ionization, the direct temperature from \oiii\ is used because the detection is coming from an ionization zone that is dominant in the region. If the low-ionization zone is dominant, then a direct \oiii\ temperature is deprioritized in favor of a \siii\ or \nii\ temperature. There are two potential issues with this approach that can manifest as discrepancies in the abundances. First, the updated T$_e$-T$_e$ relations now include larger uncertainties (see Equation \ref{eq:dT_eq}). In a region characterized by low ionization emission, applying the T$_e$\oiii\/-T$_e$\siii\ relation instead of using the direct \oiii\ temperature will set the minimum uncertainty on the inferred temperature to be $\sim$ 800 K. Such an increase in the uncertainty will result in an abundance with high uncertainty, which will deweight that region in abundance gradient determinations. The second issue is that this prioritization arrangement can reject high S/N \oiii$\lambda$4363 measurements in favor of more uncertain \siii\ or \nii\ detections. Not only will the choice of poorly measured lines further affect the uncertainties on the resulting abundances, but it could result in large, non-physical scatter in the abundances.

\renewcommand{\thefigure}{A.1}
\begin{figure}[b]
   \epsscale{0.53}
   \centering
   \plotone{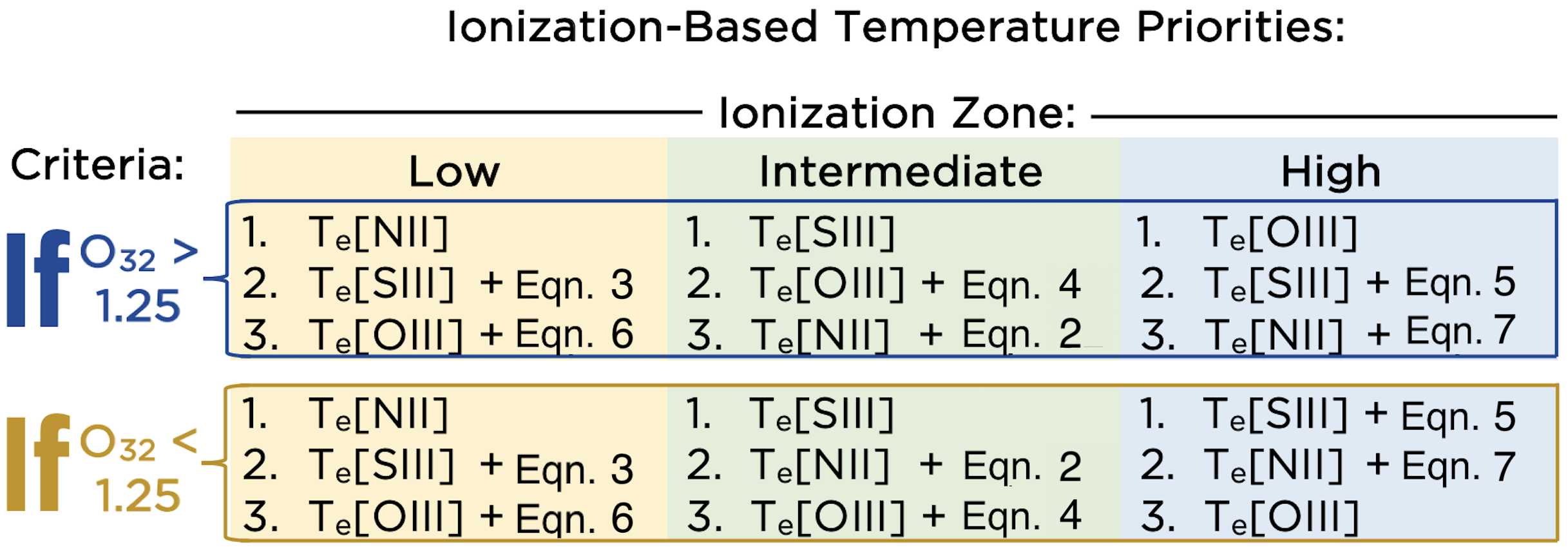}
   \caption{The ionization-based temperature prioritizations adopted by \citet{berg2019} and updated to include the equations listed in \S3.2. This designates which temperatures are to be used first (if available) for each ionization zone. The $O_{32}$ value used to decide if the temperatures are calculated for a high or low ionization region is 1.25. This Figure is modified from Figure 5 in \citet{berg2019}}
   \label{fig:temp_priors}
\end{figure}

\renewcommand{\thefigure}{A.2}
\begin{figure}
   \epsscale{0.8}
   \centering
   \plotone{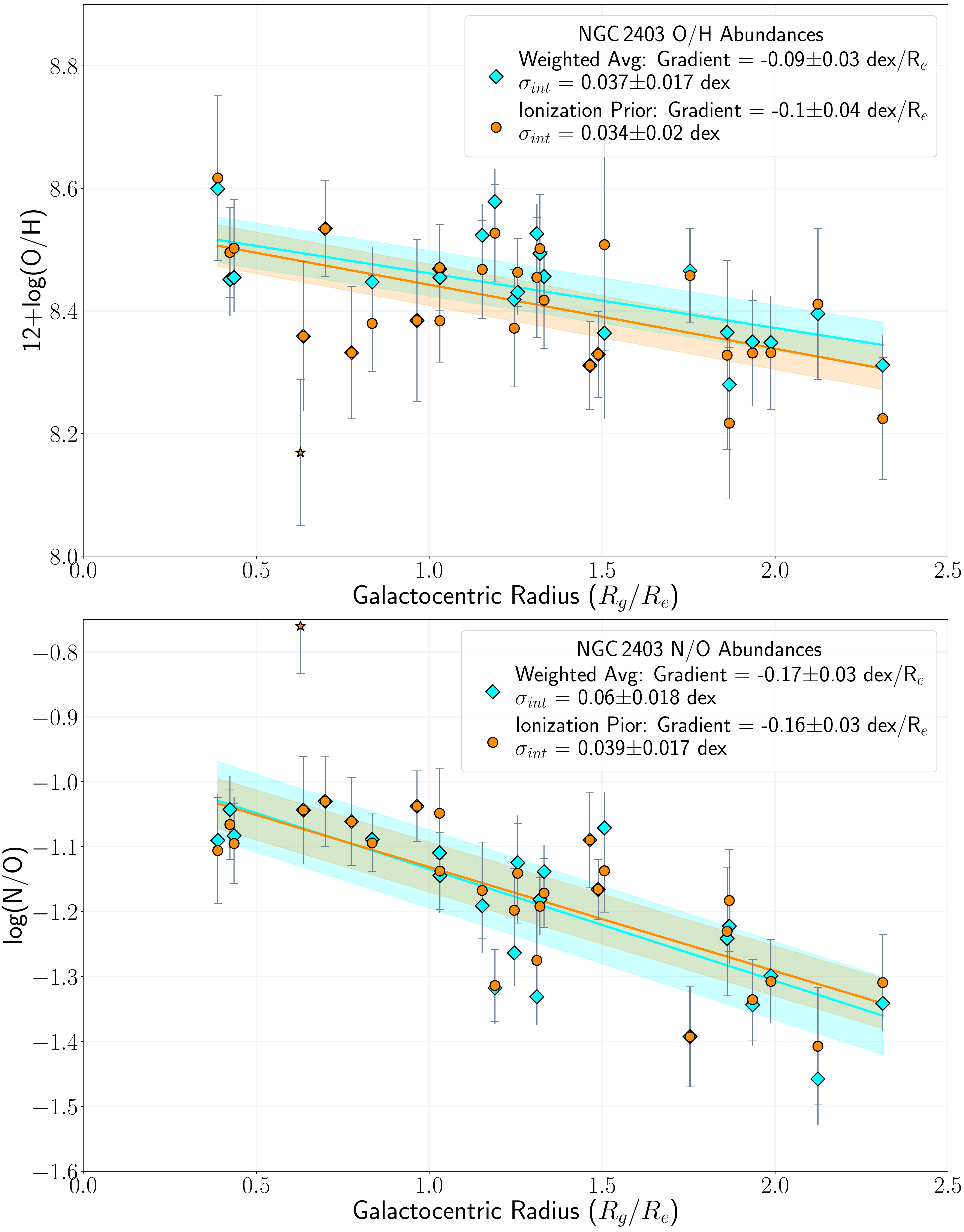}
   \caption{\textit{Top Panel}: CHAOS oxygen abundances in NGC\,2403 plotted vs.\ R$_g$/R$_e$. Cyan diamonds are abundances obtained using the weighted average ionization zone temperatures, while orange circles are the same H\ii\ regions using the ionization-based temperature prioritizations of Figure \ref{fig:temp_priors}. The capped errorbars are the errors on the abundances resulting from the ionization-based method. The gradients and intrinsic dispersions are shown in the legend of the plot. The intrinsic dispersions about each gradient are represented as the shaded region around the gradients. \textit{Bottom Panel}: Same as top panel but with N/O abundances.}
   \label{fig:n2403_IPAbuns}
\end{figure}

The possible shortcomings of the ionization-based prioritizations necessitate a comparison to other temperature prioritizations. The weighted average approach in each ionization zone utilizes the complete temperature information within an H\ii\ region. Unlike the ionization-based prioritization, this method is not as susceptible to low S/N detections unless there is a single temperature-sensitive auroral line detected in the region. Alternatively, a weighted average can be biased to the strongest auroral line detection and relies on the accuracy of the T$_e$-T$_e$ relations applied.

Figure \ref{fig:n2403_IPAbuns} plots the O/H and N/O abundances measured in the H\ii\ regions of NGC\,2403 vs.\ R$_g$/R$_e$ when using the weighted average method (cyan diamonds) and ionization-based prioritizations (orange circles) described in Figure \ref{fig:temp_priors}. Additionally, the best-fit parameters for the O/H and N/O gradients in all CHAOS galaxies when using the ionization-based prioritizations are provided in Table \ref{t:ipFits}. All but three H\ii\ regions in NGC\,2403 have O$_{32} <$ 1.25, so the majority of the regions have T$_e$\oiii\ deprioritized in the high-ionization zone. From the O/H abundances, one can see that the uncertainties are significantly larger when using the ionization-based method due to the use of T$_e$-T$_e$ relations to infer the high-ionization zone temperature. There are instances where only one auroral line is detected, in which case the O/H abundances and uncertainties as calculated by the ionization prioritization and the weighted average methods are the same. That being said, the oxygen abundances as calculated by the two different methods are consistent within each H\ii\ region in NGC\,2403.

Table \ref{t:ipCompare} provides the difference between the best-fit parameters obtained via the weighted average approach (from Table \ref{t:abunFits}) and the parameters from the ionization-based method (from Table \ref{t:ipFits}). The differences are all consistent with zero within statistical uncertainty, including the intrinsic dispersions about the gradients. The reason for this is a combination of two factors. First, while the temperatures obtained via the ionization-based prioritizations are prone to the issues discussed above, these temperatures also have large uncertainties. The large uncertainties on the resulting abundances yield intrinsic dispersions about the gradient that are generally smaller. The second factor is the small oxygen abundance uncertainty produced by the weighted average approach. Even if the total scatter is smaller about the gradient (see difference in $\sigma_{tot.}$ in Table \ref{t:ipCompare}), the reduced uncertainties result in more of the scatter being attributed to intrinsic variations rather than observational uncertainties. Given the agreement between the two methods, we conclude that the weighted average approach is at least consistent with the approach used in \citet{berg2019} and could be used as an alternative approach to determining the electron temperature within each ionization zone.

% Abundance Gradient Comparison Table
\renewcommand{\thetable}{A.3}
\begin{deluxetable*}{llccccc}[H]
\tablecaption{Ionization-Based Temperature Prioritization Abundance Fits}
\tablehead{
\CH{$y$}  & \CH{$x$}  & \CH{Galaxy} & \CH{$\#$ Reg.} & \CH{Equation} & \CH{$\sigma_{\rm{int.}}$} & \CH{$\sigma_{\rm{tot.}}$}}
\startdata
12+log(O/H) (dex)&$R_g$ ($R_e^{-1}$)  &{NGC~0628} & 45 & $y=(8.62\pm0.05) - (0.08\pm0.03)\times x$ & $0.091\pm0.018$ & 0.111 \\
                &                     & {NGC~5194} & 28 & $y=(8.59\pm0.10) - (0.02\pm0.05)\times x$ & $0.05\pm0.03$ & 0.08 \\
                &                     & {NGC~5457} & 71 & $y=(8.71\pm0.04) - (0.179\pm0.017)\times x$ & $0.087\pm0.013$ & 0.108 \\
                &                     & {NGC~3184} & 30 & $y=(8.57\pm0.17) - (0.09\pm0.12)\times x$ & $0.08\pm0.04$ & 0.14 \\
                &                     & {NGC~2403} & 27 & $y=(8.55\pm0.05) - (0.10\pm0.04)\times x$ & $0.03\pm0.02$ & 0.069 \\
\vspace{-1ex} \\             
log(N/O) (dex)  & $R_g$ ($R_e^{-1}$)  & {NGC~0628} & 45 & $y=(-0.48\pm0.04) - (0.29\pm0.02)\times x$ & $0.076\pm0.011$	& 0.081	\\
                &                     & {NGC~5194} & 28 & $y=(-0.31\pm0.06) - (0.14\pm0.03)\times x$ & $0.075\pm0.016$	& 0.079	\\
                &                     & {NGC~5457} & 71 & $y=(-0.72\pm0.03) - (0.186\pm0.016)\times x$ & $0.100\pm0.012$	& 0.112	\\
                &                     & {NGC~3184} & 30 & $y=(-0.32\pm0.09) - (0.36\pm0.06)\times x$ & $0.059\pm0.019$	& 0.081  \\
                &                     & {NGC~2403} & 27 & $y=(-0.97\pm0.04) - (0.16\pm0.03)\times x$ & $0.039\pm0.017$	& 0.066
\enddata
\tablecomments{
Best-fit equations for the O/H and N/O abundance gradients observed in the CHAOS galaxies when using ionization-based temperature prioritizations. The first and second columns are the dependent and independent variables, respectively, used for the Equation listed in the fifth column. The galaxy and the number of regions used for the fit are given in the third and fourth column. The intrinsic and total scatter, both in dex, are listed in the sixth and seventh columns, respectively.}
\label{t:ipFits}
\end{deluxetable*}

\renewcommand{\thetable}{A.4}
% Abundance Gradient Difference Table
\begin{deluxetable*}{ccccc}[H]
\tablecaption{Difference between the Weighted Average and Ionization-Based Temperature Prioritization Gradient Parameters}
\tablehead{
\CH{Galaxy} & \CH{$\Delta$ Central O/H} & \CH{$\Delta$ Gradient O/H} & \CH{$\Delta\sigma_{int}$ Around O/H} & \CH{$\Delta\sigma_{tot.}$ Around O/H}}
\startdata
{NGC~0628} & $0.02\pm0.07$ & $-0.00\pm0.04$ & $-0.01\pm0.02$ & $-0.012$ \\
{NGC~5194} & $-0.11\pm0.14$ & $0.04\pm0.07$ & $-0.01\pm0.03$ & $-0.014$ \\
{NGC~5457} & $-0.02\pm0.05$ & $0.01\pm0.02$ & $0.010\pm0.018$ & $0.005$ \\
{NGC~3184} & $0.1\pm0.2$ & $-0.05\pm0.16$ & $0.01\pm0.05$ & $-0.023$ \\
{NGC~2403} & $0.00\pm0.06$ & $0.01\pm0.05$ & $0.01\pm0.03$ & $-0.006$ \\
\vspace{-1ex} \\             
\hline\hline
\vspace{-1ex} \\ 
\CH{Galaxy} & \CH{$\Delta$ Central N/O} & \CH{$\Delta$ Gradient N/O} & \CH{$\Delta\sigma_{int}$ About N/O} & \CH{$\Delta\sigma_{tot.}$ About N/O} \\
\hline
\vspace{-1ex} \\    
{NGC~0628} & $-0.01\pm0.05$ & $0.00\pm0.03$ & $-0.004\pm0.015$	& $-0.006$	\\
{NGC~5194} & $0.03\pm0.09$ & $-0.01\pm0.05$ & $-0.003\pm0.019$	& $-0.004$	\\
{NGC~5457} & $0.02\pm0.05$ & $-0.01\pm0.02$ & $0.010\pm0.017$	& $0.008$ \\
{NGC~3184} & $-0.02\pm0.12$ & $0.01\pm0.08$ & $0.00\pm0.03$	& $-0.005$  \\
{NGC~2403} & $0.01\pm0.06$ & $-0.01\pm0.04$ & $0.021\pm0.025$	& $0.009$
\enddata
\tablecomments{
The difference between the best-fit parameters provided in Table \ref{t:abunFits} for the weighted average temperature prioritizations and in Table \ref{t:ipFits} for the ionization-based temperature prioritizations. The name of the galaxy is given in the first column; the second column provides the difference in the central abundance values; the third provides the difference between the gradients; the difference between the intrinsic and total dispersion about the gradient are provided in the fourth and fifth columns, respectively. All units are in dex. The first five rows provide the differences between the oxygen abundance gradients, while the last five rows are the differences between the N/O gradients.}
\label{t:ipCompare}
\end{deluxetable*}

\newpage

%%%%%% BIBLIOGRAPHY %%%%%%

%\bibliographystyle{apj}
%\addbibresource{test.bib}
%\printbibliography
\bibliography{chaosvi_biblio.bib}

\end{document}